\documentclass[twocolumn]{aastex631}

\usepackage{graphicx}
\usepackage{amsmath}
\usepackage{hyperref}
\usepackage{booktabs}
\usepackage{multirow}

\shorttitle{Validation and Characterization of Seven TESS Exoplanets}
\shortauthors{Escolà-Rodrigo}

\begin{document}

\title{Validation and Physical Characterization of Seven TESS Exoplanets: A Compact Multi-Planet System and Targets for Atmospheric Spectroscopy}

\author[0009-0007-1479-2386]{Biel Escolà-Rodrigo}
\affiliation{Independent Researcher}
\email{escola.rodrigo.astro@gmail.com}

\begin{abstract}
The Transiting Exoplanet Survey Satellite (TESS) continues to identify thousands of transit candidates that require statistical validation to rule out astrophysical false positives before atmospheric characterization can be pursued. We statistically validate and characterize seven TESS exoplanet candidates orbiting five host stars. As a consistency check of our automated vetting and Bayesian inference pipeline, we include TOI-2392.03, a sub-Neptune companion previously validated in the literature. Our framework combines automated signal retrieval, pixel-level centroid and Gaia DR3 contamination analysis, and false positive probability (FPP) computation with the \texttt{TRICERATOPS} vetting engine. Physical and orbital parameters are derived via nested sampling (\texttt{juliet}/Dynesty), complemented by an empirical mass-radius relation, since no dynamical mass measurements are available. Applying the same procedure to all candidates, we statistically validate six new exoplanets ($\mathrm{FPP} < 0.015$ and $\mathrm{NFPP} < 0.001$), one of which (TOI-7721.01) lies close to the FPP threshold. We report the period ratios of the three-planet system TOI-2392, and identify TOI-7721.01 and TOI-1277.02 as the candidates with the highest Transmission Spectroscopy Metrics (TSM) in our sample, making them targets of interest for atmospheric follow-up with the \textit{James Webb Space Telescope} (JWST) and the upcoming \textit{Ariel} mission.
\end{abstract}


\section{Introduction}

The Transiting Exoplanet Survey Satellite \citep[TESS;][]{Ricker2015} has transformed exoplanetary science by surveying the brightest stars across the celestial sphere. This all-sky photometric monitoring has resulted in the identification of thousands of TESS Objects of Interest (TOIs), providing a large catalog of planetary candidates \citep[e.g.,][]{Guerrero2021}. However, the volume and varied nature of these transit-like signals present an analytical challenge. A significant fraction of these detections are astrophysical false positives (FPs)---predominantly eclipsing binaries (EBs), hierarchical triples, or background eclipsing binaries (BEBs) blended within the large $\sim21\arcsec$ TESS pixels. Consequently, statistical validation is a necessary step to elevate a candidate to the status of a validated exoplanet, particularly before allocating resources for radial velocity (RV) mass determination or atmospheric spectroscopy. Building upon the probabilistic methodologies established by \citet{Torres2011} and \citet{Morton2012}, modern Bayesian engines allow these observational degeneracies to be addressed quantitatively.

Compact multi-planet systems, particularly those harboring sub-Neptunes and super-Earths, offer laboratories for understanding planetary formation, orbital migration, and dynamical stability \citep{Lissauer2012}. Furthermore, companions in temperate or highly irradiated regimes around bright host stars are of interest for atmospheric characterization with missions such as the \textit{James Webb Space Telescope} (JWST) and the ESA \textit{Ariel} mission.

In this study, we present the validation and physical characterization of seven TESS candidates spanning a range of physical regimes. Several of them have a complicated follow-up history. For instance, ground-based follow-up efforts recorded on the Exoplanet Follow-up Observation Program (ExoFOP) database report difficulty in recovering the transit of TOI-1277.02 with sufficient per-event significance, often attributing tentative detections to instrumental systematics or stellar variability---a possibility consistent with the high spot coverage and young age of the host star reported by \citet{Savanov2022}. Similarly, TOI-1737.01 and TOI-4450.01 have undergone high-resolution speckle imaging and reconnaissance spectroscopy that, while ruling out obvious stellar companions, left their planetary nature unconfirmed. Recent catalogs produced with the RAVEN machine-learning pipeline \citep{Hadjigeorghiou2025,Lafarga2026} list TOI-7721.01 and TOI-4450.01 as high-probability planet candidates, as recorded on ExoFOP; to our knowledge, no statistical validation of these candidates with \texttt{TRICERATOPS} has been published.

As a consistency check of our methodology, we include TOI-2392.03---a sub-Neptune companion recently validated with \texttt{TRICERATOPS} by \citet{Collier2026} and now cataloged as TOI-2392 b. We process it through the same vetting, pixel-level, and Bayesian nested sampling pipeline as every other candidate, without using its published parameters as input. We then compare our results with the published ones. Recovering the published parameters and FPP of one planet checks the parameter extraction and gives one independent FPP comparison, but does not by itself calibrate the false-positive rejection power of the pipeline; the validation of the remaining candidates rests on the \texttt{TRICERATOPS} probabilities and the supporting checks described below.

This paper is structured as follows: In Section 2, we outline the target selection and vetting funnel that led to our seven candidates. Section 3 details the observational data, including both pre-conditioned and raw TESS photometry. In Section 4, we perform the false-positive vetting, applying Box Least Squares (BLS) signal retrieval, Gaia DR3 contamination constraints, and computing the False Positive Probabilities (FPP) with \texttt{TRICERATOPS} \citep{Giacalone2021}. Section 5 describes the physical parameter derivation through nested sampling with \texttt{juliet} \citep{Espinoza2019}. Finally, Sections 6 and 7 present the validated systems, place them in the context of the radius valley, and assess their potential for transmission spectroscopy.

\section{Target Selection and Sample Reduction}

The transition from a raw list of transit-like threshold crossing events (TCEs) to a reliable sample of planetary candidates requires a systematic selection process. Our strategy was designed to isolate signals with high signal-to-noise ratios (SNR) around relatively bright stars, thereby favoring the feasibility of subsequent precision radial velocity (PRV) mass measurements and transmission spectroscopy.

\subsection{Initial Sample Definition}
Our initial sample was sourced from the publicly available NASA Exoplanet Archive TOI catalog. We applied a series of photometric and physical thresholds based on the community parameters reported on ExoFOP. Specifically, we required a minimum cataloged transit depth of 200 parts-per-million (ppm) and an aggregate SNR $\geq 20$ on ExoFOP, so that the preliminary transits were distinguishable from instrumental red noise (our subsequent custom phase-masking SNR evaluations on the un-detrended PDCSAP and SAP light curves can yield lower values than the cataloged ExoFOP SNR). To facilitate potential ground-based follow-up, we restricted the host star apparent brightness to $T_{\mathrm{mag}} < 13$. Furthermore, to focus the search on the terrestrial-to-Jovian regime and exclude obvious stellar eclipsing binaries, we constrained the preliminary estimated planetary radii to $0.5 < R_p < 8.0\,R_\oplus$. This primary filtering yielded an initial sample of approximately 228 candidate systems.

\subsection{The Vetting Funnel}
Despite these criteria, the TESS catalog contains a high fraction of astrophysical false positives, particularly unresolved bound stellar companions or dense background blends within the large TESS photometric apertures. To reduce the sample, we employed a multi-step vetting funnel.

First, we cross-matched the host stars with the Gaia Data Release 3 \citep[DR3;][]{Gaia2023} catalog to extract the Renormalized Unit Weight Error (RUWE). The RUWE metric is an indicator of unmodeled astrometric perturbation, often associated with an unresolved massive binary companion. We discarded any target with RUWE $> 1.4$, so that our host stars are astrometrically well-behaved single stars (or wide, non-interfering binaries).

Second, we reviewed the community-contributed observational notes and high-resolution imaging data housed on the ExoFOP platform. Candidates flagged as clear background eclipsing binaries (BEBs) or exhibiting nearby bright stellar contaminants within the TESS aperture that could not be resolved were removed. This astrometric and community-vetting step reduced the pool to roughly 95 candidates.

At this stage, we excluded candidates that already had strong independent confirmation evidence on ExoFOP or were explicitly flagged as being in active preparation for publication by other teams, to avoid duplicating ongoing work. The remaining candidates were run through our BLS search. A large fraction was discarded at this step, because the transit signal was not recovered or was recovered only at low SNR owing to the photometric noise of the host star. No fixed SNR threshold was imposed at this step; all the planets presented here were recovered with SNR above 10 (Table \ref{tab:signals}). The candidates whose signal was recovered were then passed to the pixel-level vetting and the \texttt{TRICERATOPS} analysis; those that do not appear in this paper did not satisfy the validation thresholds ($\mathrm{FPP} < 0.015$ and $\mathrm{NFPP} < 0.001$). The seven planets presented here are those that satisfied the thresholds, so our sample is the set of candidates that passed this pipeline and not a complete or unbiased survey. The number of candidates discarded at each stage was not tabulated. They span a range of physical regimes---from highly irradiated sub-Saturns to a compact multi-planet system---and several present observational challenges that test the pipeline.

\section{Observations and Stellar Parameters}

Our validation framework relies on high-cadence space-based photometry combined with stellar characterization from public catalogs.

\subsection{TESS Photometry: PDCSAP and SAP Fluxes}
For all seven targets, we queried the Mikulski Archive for Space Telescopes (MAST) using the \texttt{Lightkurve} Python package \citep{Lightkurve2018} to retrieve the 2-minute cadence time-series photometry processed by the Science Processing Operations Center (SPOC). For each target we used all available sectors with 2-minute cadence (Table \ref{tab:signals}). We extracted both the Pre-search Data Conditioning Simple Aperture Photometry (PDCSAP) and the raw Simple Aperture Photometry (SAP).

The PDCSAP flux, which uses co-trending basis vectors to remove common instrumental systematics, was used for the period search and the final Bayesian transit modeling. The raw SAP flux, alongside the corresponding Target Pixel Files (TPFs), was used for spatial contamination analyses and FPP evaluations. Access to the SAP flux is needed because PDCSAP routines automatically apply crowding corrections that modify the dilution fraction from background stars, which is a required input for the astrophysical false positive probability calculation.

\subsection{Host Star Properties}
The inferred planetary properties depend directly on the accuracy of the host star parameters. We adopted the stellar effective temperatures ($T_{\mathrm{eff}}$), masses ($M_*$), radii ($R_*$), and surface gravities ($\log g$) from the TESS Input Catalog \citep[TIC v8;][]{Stassun2019}, cross-checked with Gaia DR3 \citep{Gaia2023}. The compiled stellar parameters, alongside their Gaia RUWE statistics, are presented in Table \ref{tab:stellar_params}. The table lists five host stars, since TOI-2392 hosts three of the seven planets.
For TOI-7721, which lacks reconnaissance spectroscopy in the archival database, we report an assumed solar metallicity ([Fe/H] = 0.0) with an uncertainty of $\pm 0.5$ dex for reference. The limb-darkening priors were assigned independently of stellar metallicity, as described in Section 5.1.
\begin{table*}[ht]
\centering
\caption{Host Star Properties and Astrometric Constraints}
\label{tab:stellar_params}
\begin{tabular}{lccccc}
\toprule
\textbf{Target (TIC)} & \textbf{$T_{\mathrm{eff}}$ (K)} & \textbf{$M_*$ ($M_\odot$)} & \textbf{$R_*$ ($R_\odot$)} & \textbf{$\log g$ (cgs)} & \textbf{RUWE} \\
\midrule
TOI-1277 (153949511) & $5544 \pm 120$ & $0.98 \pm 0.12$ & $0.86 \pm 0.04$ & $4.56$ & $0.85$ \\
TOI-1737 (233009109) & $6669 \pm 128$ & $1.41 \pm 0.24$ & $1.95 \pm 0.08$ & $4.01$ & $0.96$ \\
TOI-2392 (149349867) & $6244 \pm 134$ & $1.21 \pm 0.18$ & $1.76 \pm 0.09$ & $4.03$ & $0.81$ \\
TOI-4450 (373017346) & $5844 \pm 135$ & $1.05 \pm 0.13$ & $1.15 \pm 0.06$ & $4.34$ & $0.76$ \\
TOI-7721 (163029327) & $5415 \pm 126$ & $0.94 \pm 0.11$ & $0.95 \pm 0.05$ & $4.46$ & $0.94$ \\
\bottomrule
\end{tabular}
\end{table*}

\section{False Positive Vetting and Statistical Validation}

Before planetary parameters are derived, we quantitatively assess astrophysical false positive scenarios. Our vetting pipeline has three steps: signal retrieval, pixel-level centroid tracking combined with spatial contamination analysis, and Bayesian probability computation.

\subsection{Signal Retrieval and Morphological Analysis}

To refine the ephemerides and obtain consistent period constraints across all candidates, we developed a dual-track automated detection pipeline based on \texttt{Lightkurve}.

In the primary track, we retrieved the PDCSAP flux for all available TESS sectors. Each sector was individually normalized and cleaned with an asymmetric sigma-clipping routine before stitching the multi-sector data. We applied a $4\sigma$ upper limit to remove stellar flares and cosmic ray hits, and a conservative $20\sigma$ lower limit to preserve deep transit events. To improve the detectability of shallow transits against residual stellar variability and instrumental red noise, the PDCSAP light curves were flattened with a Savitzky-Golay filter whose window was set to at least three times the maximum expected transit duration (with a lower limit of 11 cadences), to avoid suppressing the transit signal. The flattened flux was then searched with a Box Least Squares (BLS) periodogram \citep{Kovacs2002} over a finely sampled period grid to recover the period ($P$), mid-transit epoch ($T_0$), and transit duration ($T_{14}$). Once the ephemerides were identified, the flattened data were not used further.

Returning to the non-flattened PDCSAP flux, we performed a visual odd/even transit comparison by separately phase-folding the odd- and even-numbered transits. We observed no apparent differences in transit depth or duration between alternating events, which disfavors the scenario of a nearly equal-mass eclipsing binary orbiting at twice the reported period.

In parallel, the secondary track extracted the raw SAP flux for all targets. Because PDCSAP routines apply crowding corrections that alter the raw photometric dilution needed for the false positive calculation, the uncorrected SAP data are required for this step. We applied an asymmetric outlier rejection ($4\sigma$ upper and $40\sigma$ lower limits) to the SAP flux and phase-folded the data using the ephemerides derived from an identical BLS search on the flattened SAP light curve. The un-detrended and unbinned phase-folded SAP light curves were then exported as the input data for the Bayesian false positive evaluation.

To compute the transit depth and the signal-to-noise ratio (SNR), we applied an identical phase-masking strategy independently to the un-detrended PDCSAP and SAP light curves. The transit depth was derived from the central 50\% of the transit duration (within $\pm 0.25\,T_{14}$ of mid-transit) to avoid distortions during ingress and egress, while the out-of-transit baseline, used to establish the empirical noise variance ($\sigma$), was sampled symmetrically from windows outside the transit (between 0.75 and 2.25 times the transit duration from mid-transit).

The observational sectors, transit durations, PDCSAP transit depths, and the SNR values measured independently with this method on both the PDCSAP and SAP light curves are summarized in Table \ref{tab:signals}.

\begin{table*}[ht]
\centering
\caption{TESS Observations and Measured Signal Properties}
\label{tab:signals}
\resizebox{\textwidth}{!}{%
\begin{tabular}{lcccccc}
\toprule
\textbf{Candidate} & \textbf{Sectors (2-min SPOC)} & \textbf{$N_{\mathrm{sectors}}$} & \textbf{$T_{14}$ (h)} & \textbf{Depth (ppm)} & \textbf{SNR (PDCSAP)} & \textbf{SNR (SAP)} \\
\midrule
TOI-1277.02 & 14--15, 21, 47--48, 74--75 & 7 & 3.84 & 1287 & 43.90 & 14.18 \\
TOI-1737.01 & 15--26, 40--41, 47--60, 73--78, 80, 82--86 & 40 & 5.76 & 245 & 50.01 & 20.63 \\
TOI-2392.01 & 27--30, 32--33, 35--39, 61, 63--64, 66--69, 87--88, 90, 93, 95--98 & 26 & 4.44 & 410 & 37.77 & 29.98 \\
TOI-2392.02 & 27--30, 32--33, 35--39, 61, 63--64, 66--69, 87--88, 90, 93, 95--98 & 26 & 3.84 & 284 & 31.87 & 27.51 \\
TOI-2392.03 & 27--30, 32--33, 35--39, 61, 63--64, 66--69, 87--88, 90, 93, 95--98 & 26 & 5.16 & 216 & 18.23 & 11.99 \\
TOI-4450.01 & 40, 47, 52--53, 59--60, 79, 86 & 8 & 3.12 & 1224 & 35.60 & 27.47 \\
TOI-7721.01 & 99--101 & 3 & 1.68 & 5537 & 44.41 & 28.80 \\
\bottomrule
\multicolumn{7}{l}{\footnotesize \textit{Note:} Transit duration is from the BLS search; depth is measured on the un-detrended PDCSAP light curves.} \\
\multicolumn{7}{l}{\footnotesize The two SNR values are calculated separately for the PDCSAP and SAP light curves using the same phase-masking method.} \\
\end{tabular}%
}
\end{table*}

\subsection{Pixel-Level Astrometry and Spatial Contamination Analysis}

The large pixel scale of TESS ($\sim21\arcsec$/pixel) merges the flux of the target star with nearby sources, making transit signals susceptible to blending from background eclipsing binaries (BEBs). To assess these scenarios prior to the Bayesian probability evaluation, we performed a joint astrometric and photometric centroid analysis.

First, to evaluate spatial contamination, we developed an automated multi-sector retrieval script interfacing with the Gaia DR3 catalog \citep{Gaia2023}. For every available TESS sector (including full-frame image \texttt{TESS-SPOC} sectors), the algorithm cross-matches the astrometric coordinates and $G$-band magnitudes of all Gaia sources within a $1\arcmin$ radius against the SPOC photometric pipeline mask applied for that sector, correcting for stellar proper motions. Rather than adopting an average contamination metric, our pipeline identifies the ``worst-case'' scenario, i.e., the sector with the highest photometric dilution. This maximum flux dilution percentage is calculated as the ratio of the integrated flux of all neighboring sources falling within the SPOC mask to the flux of the target star, derived from their Gaia $G$-band magnitudes ($F \propto 10^{-0.4 G_{\mathrm{mag}}}$). We note that this dilution is computed in the Gaia $G$ band and not in the TESS band, so it may be inaccurate for neighbors whose colors differ substantially from those of the target.

\begin{figure*}[ht]
    \centering
    \includegraphics[width=0.6\textwidth]{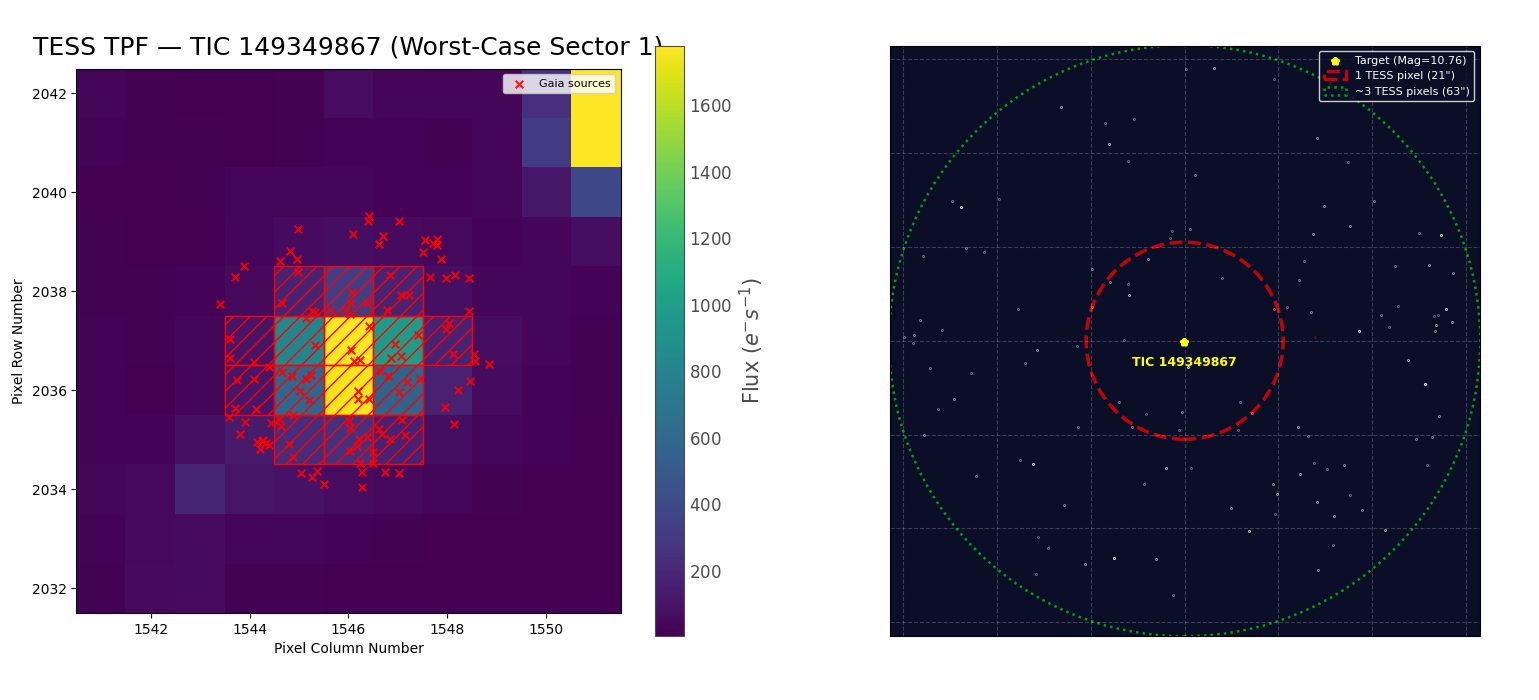}
    \caption{Spatial contamination analysis for TOI-2392.03 (worst-case scenario, Sector 1). \textit{Left}: TESS Target Pixel File overlaid with the SPOC pipeline aperture mask. \textit{Right}: Gaia DR3 map ($1\arcmin$ search radius) showing proper-motion-corrected neighboring sources. Despite the high density of background stars, their combined flux contribution inside the SPOC mask is small.}
    \label{fig:sc_2392}
    \vspace*{1cm}
\end{figure*}

For TOI-2392.03, visual inspection of the Target Pixel Files suggests a densely populated background (see Figure \ref{fig:sc_2392}), but our Gaia DR3 flux integration shows that the neighboring sources are faint, contributing a small fraction (3.61\%) of the total flux within the aperture in the worst-case sector. For all seven candidates, we report the worst-case flux dilution percentage together with the number of neighboring sources within the search radius in Table \ref{tab:validation}. The contamination maps and Gaia DR3 overlays for the remaining systems are provided in Appendix A for completeness.

Second, we analyzed the sub-pixel photometric centroid stability during the transit events. Using the raw Target Pixel Files (TPFs) from the most recent observational sector, we extracted the flux-weighted moment centroids. To isolate a possible centroid shift caused by an off-target eclipse, we removed the spacecraft's pointing drift with a median filter whose kernel size was set to exceed three times the transit duration, to avoid self-subtraction of the centroid signal.

The drift-corrected centroid time series along the column ($\Delta X$) and row ($\Delta Y$) axes were phase-folded to the BLS ephemerides and binned at 30-minute resolution. For all seven candidates, the maximum in-transit centroid displacement is below 0.005 pixels, a nominal threshold that we adopt for this check; we therefore find no displacement above this threshold. We did not estimate formal uncertainties for these displacements, so we regard this test as a coarse, complementary check: the in-transit displacement expected from a blended source depends on its separation and the transit depth, and the weight of the validation is carried by the \texttt{TRICERATOPS} analysis described below. The worst-case flux dilution and maximum centroid displacements are reported in Table \ref{tab:validation} alongside the Bayesian validation probabilities.

\subsection{Bayesian False Positive Probabilities (\texttt{TRICERATOPS})}

To quantify the probability that each signal is a planet, we used the \texttt{TRICERATOPS} vetting engine \citep{Giacalone2021}. This Bayesian framework computes a False Positive Probability (FPP) and a Nearby False Positive Probability (NFPP) by modeling the marginal likelihoods of various transit-producing scenarios---such as transiting planets, unresolved eclipsing binaries, or blended background eclipsing binaries---given the TESS photometry and the SPOC pipeline apertures.

For each candidate, we supplied the raw, un-detrended SAP light curve (Section 4.1) alongside the SPOC pipeline masks for the 2-minute cadence sectors listed in Table \ref{tab:signals}. The phase-folded light curves were cropped to a window adapted to each candidate's transit duration, and then locally re-normalized using a 95th-percentile out-of-transit flux constraint.
We did not use high-resolution imaging contrast curves. Such constraints generally lower the FPP, since they exclude false-positive scenarios involving unresolved companions: \citet{Collier2026} find that the FPP increases when the contrast curve is removed for 88\% of the targets that have one. Omitting them is therefore conservative, and all the candidates presented here satisfy the validation thresholds without them.

Because marginal likelihoods computed via Markov Chain Monte Carlo (MCMC) sampling are stochastic, a single evaluation can give a non-representative probability. Following the recommendations of the \texttt{TRICERATOPS} developers, we executed 20 independent \texttt{TRICERATOPS} runs per target and report the mean of these 20 iterations, without rejecting any run. A candidate is considered statistically validated if its mean probabilities across the 20 runs satisfy $\mathrm{FPP} < 0.015$ and $\mathrm{NFPP} < 0.001$ \citep{Giacalone2021,Collier2026}.

For the consistency check on TOI-2392.03, \texttt{TRICERATOPS} yielded a mean $\mathrm{FPP} = 0.00841$ and $\mathrm{NFPP} = 2 \times 10^{-5}$, satisfying the validation criteria. Applying the same procedure to the remaining candidates, all six satisfy the thresholds. Five of them do so with FPP below 0.01. TOI-7721.01 has $\mathrm{FPP} = 0.0137$, close to the 0.015 threshold, and also has the highest neighbor dilution of the sample (8.31\%); we therefore consider its validation marginal, and sensitive to the adopted inputs. The aggregated vetting metrics---spatial dilution, centroid displacements, and Bayesian probabilities---are presented in Table \ref{tab:validation}.

\begin{table*}[ht]
\centering
\caption{Statistical Validation and False Positive Vetting Metrics}
\label{tab:validation}
\resizebox{\textwidth}{!}{%
\begin{tabular}{lcccccc}
\toprule
\multirow{2}{*}{\textbf{Target (TOI)}} & \multicolumn{2}{c}{\textbf{Spatial Contamination}} & \multicolumn{2}{c}{\textbf{Centroid Shift (pixels)}} & \multicolumn{2}{c}{\textbf{TRICERATOPS (Mean)}} \\
\cmidrule(lr){2-3} \cmidrule(lr){4-5} \cmidrule(lr){6-7}
 & $N_{\mathrm{neighbors}}$ ($<1\arcmin$) & Max Neighbor-to-Target Flux (\%) & $\Delta X$ & $\Delta Y$ & FPP & NFPP \\
\midrule
TOI-1277.02 & 3 & 0.11 & 0.0006 & 0.0005 & $0.00287$ & $< 10^{-8}$ \\
TOI-1737.01 & 10 & 1.37 & 0.0004 & 0.0004 & $0.00187$ & $4 \times 10^{-7}$ \\
TOI-2392.01 & 116 & 3.61 & 0.0001 & 0.0001 & $0.00219$ & $< 10^{-8}$ \\
TOI-2392.02 & 116 & 3.61 & 0.0005 & 0.0009 & $0.00563$ & $2 \times 10^{-7}$ \\
TOI-2392.03 & 116 & 3.61 & 0.0008 & 0.0012 & $0.00841$ & $2 \times 10^{-5}$ \\
TOI-4450.01 & 5 & 0.17 & 0.0014 & 0.0027 & $0.00823$ & $< 10^{-8}$ \\
TOI-7721.01 & 43 & 8.31 & 0.0003 & 0.0003 & $0.01371$ & $< 10^{-8}$ \\
\bottomrule
\multicolumn{7}{l}{\footnotesize \textit{Note:} $N_{\mathrm{neighbors}}$ excludes the target star. Spatial contamination metrics represent the ``worst-case'' observational sector, computed in the Gaia $G$ band.} \\
\multicolumn{7}{l}{\footnotesize Centroid shifts correspond to the drift-corrected maximum in-transit displacement.} \\
\end{tabular}%
}
\vspace*{1cm}
\end{table*}

\section{Light Curve Modeling \& Physical Parameter Derivation}

\subsection{Bayesian Inference with \texttt{juliet}}

To extract planetary parameters from the validated transit signals, we performed Bayesian modeling with the \texttt{juliet} framework \citep{Espinoza2019}. The unclipped PDCSAP TESS photometry was fit as a time series (not phase-folded): we stitched the multi-sector light curves, kept the observation times, and selected the data within $\pm 2$ transit durations of each predicted mid-transit, using the input ephemeris from our BLS search (Section 4.1), to preserve an out-of-transit baseline, and binned the light curves in 3-minute bins. Each planet of the TOI-2392 system was fit independently with a single-planet model; consequently, transits of the other two planets fall inside the fit window in roughly one quarter of the transits and were not masked, which may increase the scatter of individual transits and could introduce a small bias in the transit depths that we did not quantify.

We modeled the transits with the \texttt{batman} package \citep{Kreidberg2015} integrated within \texttt{juliet}. The parameter space was sampled with the \texttt{Dynesty} nested sampling algorithm \citep{Speagle2020}, using 1000 live points. We adopted Gaussian priors on $P$ and $T_0$ centered on the input BLS ephemerides, with fixed standard deviations of $10^{-3}$ d and $5\times 10^{-3}$ d, respectively; uniform priors on the planet-to-star radius ratio ($p = R_p/R_*$), between 0.2 and 2.5 times the square root of the input transit depth (Section 4.1); and a uniform prior on the impact parameter ($b$), bounded between 0 and 0.99 to avoid grazing geometries. The dilution factor was fixed to 1, since the PDCSAP flux is already corrected for crowding, while we included a Gaussian prior ($\mu = 0$, $\sigma = 0.1$) on the relative flux offset and a log-uniform prior ($0.1$--$10^4\,\mathrm{ppm}$) on the instrumental white-noise jitter term. Because no precision radial velocity (PRV) constraints or secondary eclipses are available, we assumed circular orbits for all candidates, fixing the eccentricity ($e$) to 0 and the argument of periastron ($\omega$) to $90^\circ$.

Instead of fitting the scaled semi-major axis ($a/R_*$) directly, we parameterized the transit using the stellar bulk density ($\rho_*$). We adopted a Gaussian prior for $\rho_*$ obtained by propagating the uncertainties of the host star's mass and radius. The stellar limb darkening was modeled with a quadratic law in the $(q_1, q_2)$ parameterization, with Gaussian priors ($\sigma = 0.05$) centered on $(q_1, q_2) = (0.348, 0.246)$ for the TESS band, the same for all targets.

For most of the sample, the transits are too shallow for useful ground-based photometry, and the modeling was driven by TESS data alone. For TOI-1277.02, whose transit is deeper ($\sim 1.3$ ppt), we also used ground-based follow-up photometry from the Las Cumbres Observatory (LCO) Teide 1.0 m telescope, obtained on 2023 May 1 in the Sloan/Pan-STARRS $zs$ (z-short) passband and covering the ingress only. The observers reported a possible ingress detection ($\sim 1.2$ ppt, $6.1\sigma$) about 30 minutes later than the predicted time, and noted that the event could be affected by systematics (ExoFOP-TESS observation notes). These data were modeled simultaneously with the TESS data, using an independent flux offset and jitter term and band-specific Gaussian limb-darkening priors ($\sigma = 0.05$, centered on $(q_1,q_2) = (0.285, 0.231)$ for the $zs$ band). No baseline trend was fit to these data. A control fit using the TESS photometry alone yields consistent parameters (e.g., $P = 14.85592 \pm 0.00002$ d, $R_p = 3.11_{-0.16}^{+0.17}\,R_\oplus$; all parameters within $1\sigma$ of the joint fit), indicating that the partial ground-based ingress does not significantly affect the solution.

\subsection{Mass Estimation \& Derived Physics}

Our candidate sample lacks dynamical mass measurements from precision radial velocity observations. We therefore used the \texttt{mr\_forecast} empirical mass-radius framework \citep{Chen2017} to infer the planetary masses. Given our posterior radius distributions, the algorithm projects a planetary mass ($M_p$) by sampling the empirical terrestrial, Neptunian, and Jovian relations. These masses are statistical projections based on the radius, not measurements. They are used as a proxy to estimate bulk properties, the atmospheric scale height, and the TSM prior to dedicated PRV follow-up. Since the mass is predicted from the radius alone, the derived densities carry no information beyond the radius, and should not be interpreted as independent constraints on composition.

To derive the bulk physical parameters, we built a 10,000-iteration Monte Carlo error propagation. For each iteration, the host star radius $R_*$ and effective temperature $T_{\mathrm{eff}}$ were sampled from Gaussian distributions based on their cataloged uncertainties, while the orbital parameters ($P$, $T_0$, $p$, $b$, $\rho_*$) were drawn jointly from the \texttt{juliet} posterior chains, preserving their correlations.

The scaled semi-major axis was derived from the $\rho_*$ posterior (clipped at a lower bound of $10\,\mathrm{kg\,m^{-3}}$), $a/R_* = (G\rho_*P^2/3\pi)^{1/3}$, and converted to $a$ using the sampled $R_*$. The planetary equilibrium temperature ($T_{\mathrm{eq}}$) was calculated assuming uniform heat redistribution and a Bond albedo of $A_B \approx 0.3$, i.e., a $(0.7)^{0.25}$ scaling of the zero-albedo value. Using the inferred mass and the fitted radius, we derived the bulk density ($\rho_p$) and surface gravity ($g_p$). To evaluate atmospheric observability, we estimated the atmospheric scale height ($H_{\mathrm{atm}}$), assigning the mean molecular weight ($\mu$) according to the planetary radius: $\mu = 29$ for $<1.5\,R_\oplus$, $\mu = 18$ for $1.5-2.75\,R_\oplus$, and $\mu = 2.3$ for $>2.75\,R_\oplus$. These assignments are assumptions, not measurements of composition.

Finally, to compare the suitability of our sample for atmospheric characterization with JWST and \textit{Ariel}, we computed the Transmission Spectroscopy Metric (TSM) following \citet{Kempton2018}, using the 2MASS $J$ magnitude of each host and evaluating it at the median values of $R_p$, $M_p$, and $T_{\mathrm{eq}}$ (it is not propagated sample by sample). The TSM scales linearly with the equilibrium temperature. \citet{Kempton2018} adopt a zero-albedo equilibrium temperature, whereas we use the $T_{\mathrm{eq}}$ of Table \ref{tab:params_master}, which assumes $A_B \approx 0.3$. Our TSM values are therefore about 8\% lower than they would be with the zero-albedo convention. Because the TSM is inversely proportional to the mass, which in our case is poorly constrained (Table \ref{tab:params_master}), the TSM values should be read as relative rankings within the sample.

\begin{figure*}[ht]
    \centering
    \includegraphics[width=0.65\textwidth]{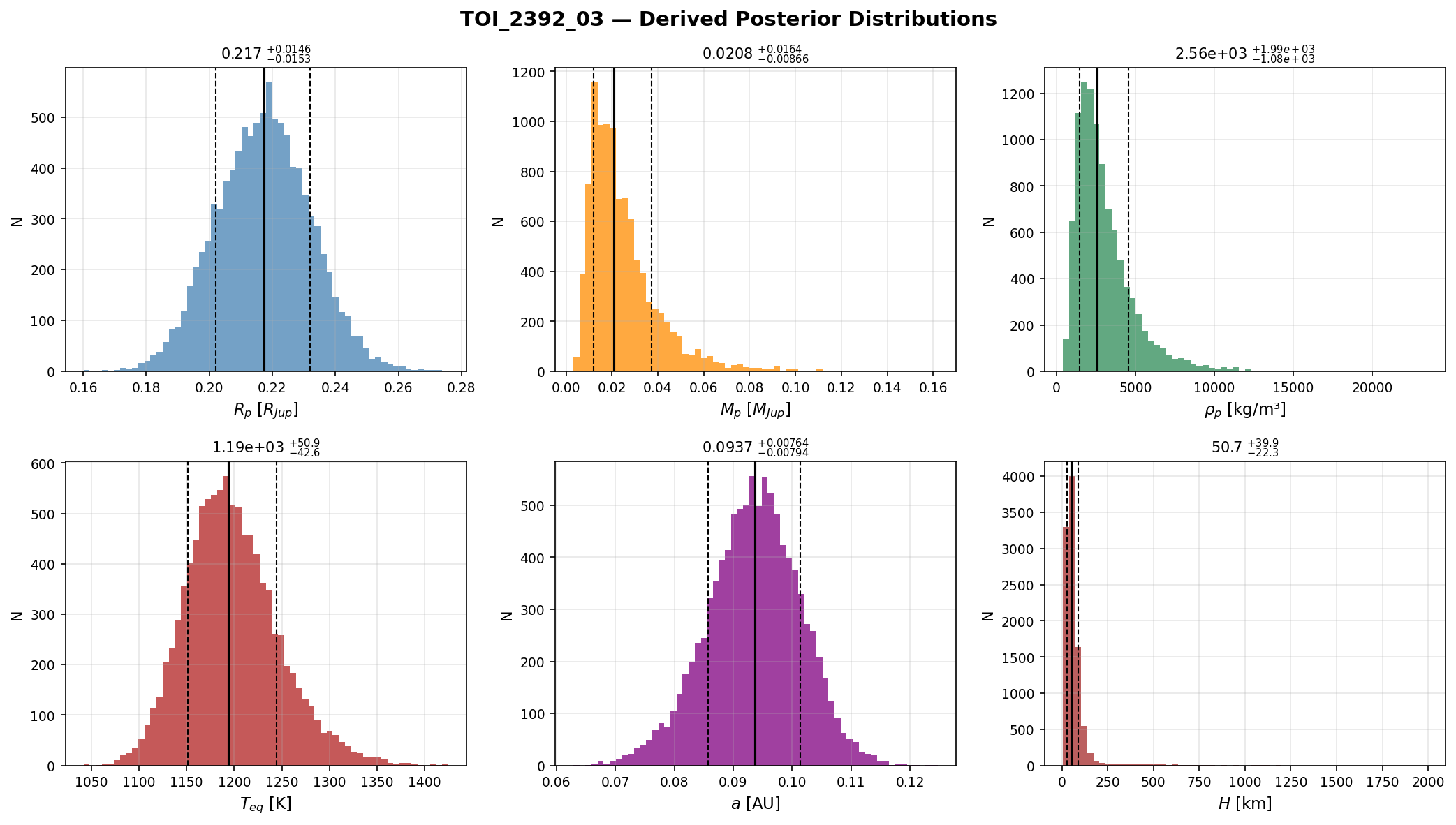}
    \caption{Derived physical posterior distributions for TOI-2392.03. The histograms show the results of the 10,000-iteration Monte Carlo propagation combining the \texttt{juliet}/\texttt{Dynesty} orbital geometry posteriors with the empirical mass-radius relation. Vertical lines denote the median (solid) and $1\sigma$ (dashed) confidence intervals.}
    \label{fig:posteriors_2392}
\end{figure*}

The resulting physical properties, orbital ephemerides, and propagated uncertainties (median and $1\sigma$ asymmetric percentiles) are given in Table \ref{tab:params_master}. Figure \ref{fig:posteriors_2392} presents the posterior distributions for TOI-2392.03. The posterior distributions for the remaining candidates are provided in Appendix B.

\section{Results \& System Architectures}

\subsection{Consistency Check: TOI-2392.03}

To test our pipeline, we processed the sub-Neptune companion TOI-2392.03 (cataloged as TOI-2392 b), previously validated by \citet{Collier2026}, with the same procedure as the other candidates. Using multi-sector TESS photometry, our retrieval and vetting steps isolated the transit signature, passed the spatial contamination checks, and satisfied the Bayesian false positive thresholds ($\mathrm{FPP} = 0.00841$).

\begin{figure}[ht]
    \centering
    \includegraphics[width=\columnwidth]{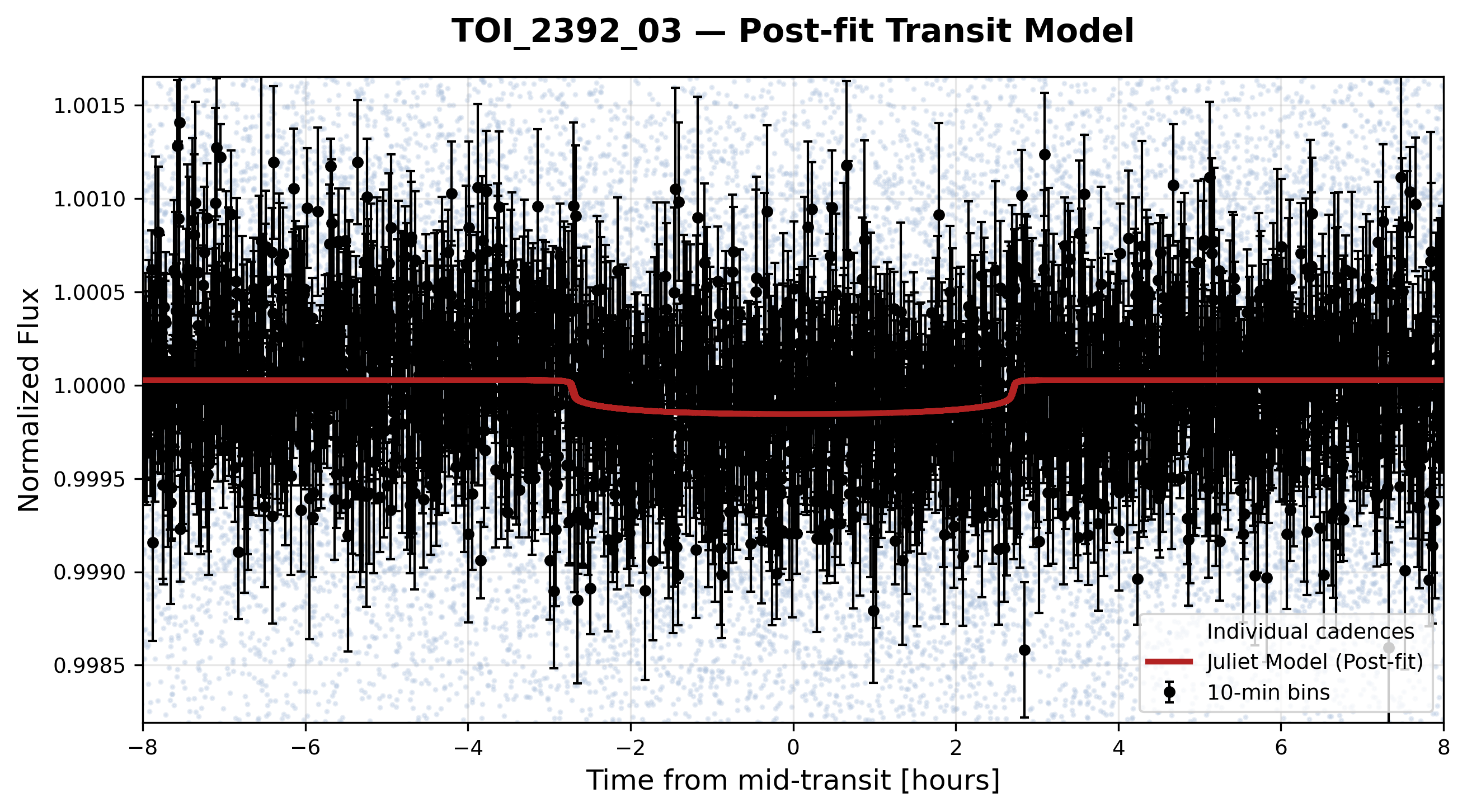}
    \caption{Phase-folded TESS photometry and Bayesian post-fit transit model for TOI-2392.03. The faint blue points represent the individual PDCSAP cadences, while the black markers with error bars denote the data binned to a 10-minute resolution for visual clarity. The solid red line represents the median \texttt{juliet} transit model. Note that the nested sampling was executed on 3-minute bins (Section 5.1); the 10-minute binning is used only for visualization.}
    \label{fig:lc_2392}
\end{figure}

The \texttt{juliet}/Dynesty modeling converged on a 9.169-day period and an impact parameter of $b = 0.45_{-0.17}^{+0.13}$. The photometric transit fit is presented in Figure \ref{fig:lc_2392}.

Our results are consistent with the published ones. We recover an orbital period of $P = 9.169453_{-0.000048}^{+0.000034}$ days, compared with the published $9.169362 \pm 0.000065$ days \citep{Collier2026}: a difference of 7.9 seconds, or $1.1\sigma$ when combining the uncertainties in quadrature. Our transit epoch, $T_0 = 2038.6294_{-0.0033}^{+0.0040}$ BTJD ($\mathrm{BJD} - 2457000$), differs from the published $2038.6340_{-0.0061}^{+0.0068}$ BTJD by $0.6\sigma$. Our FPP is also of the same order as the published one: without imaging constraints, \citet{Collier2026} report $\mathrm{FPP} = 4.9\times 10^{-3}$ for this planet ($8.9\times 10^{-4}$ with their Gemini speckle contrast curve), compared with our mean $\mathrm{FPP} = 8.41\times 10^{-3}$ obtained without high-resolution imaging.

Our Monte Carlo error propagation yields a median planetary radius of $R_p = 2.44_{-0.17}^{+0.16}\,R_\oplus$, in agreement with the published radius of $2.37 \pm 0.17\,R_\oplus$ \citep{Collier2026}; the difference is $0.3\sigma$ when combining the quoted uncertainties in quadrature. Combining this radius with the \texttt{mr\_forecast} relation, we infer a planetary mass of $M_p = 6.6_{-2.8}^{+5.2}\,M_\oplus$ and a bulk density of $\rho_p \approx 2561\,\mathrm{kg\,m^{-3}}$ under the assumption of a circular orbit. Given that the mass is predicted from the radius, this is consistent with, but does not demonstrate, a volatile-rich sub-Neptune.

With an equilibrium temperature of $T_{\mathrm{eq}} \approx 1194\,\mathrm{K}$ and an estimated atmospheric scale height ($H_{\mathrm{atm}}$) of about 51 km, TOI-2392.03 has a TSM of 11.3, the lowest of our sample. We include it as a methodological check and do not single it out as a spectroscopy target. This test shows that our parameter extraction recovers the published values for this planet; it does not by itself measure the false-positive rejection performance of the pipeline.

\subsection{The Newly Validated Planets}

\begin{figure*}[ht]
    \centering
    \includegraphics[width=0.75\textwidth]{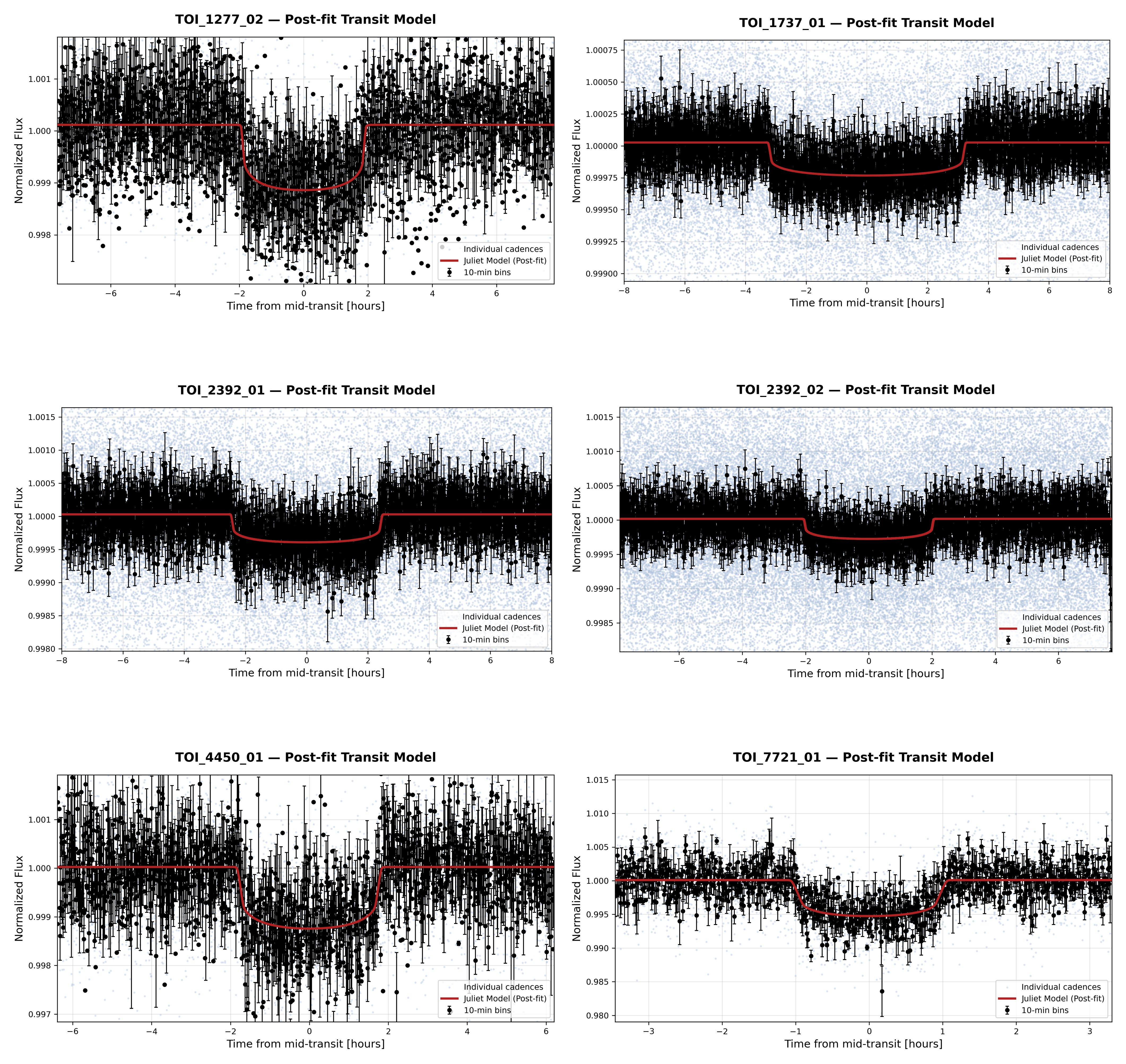}
    \caption{Phase-folded TESS photometry and \texttt{juliet} transit models for the six newly validated exoplanets. The individual PDCSAP cadences (faint blue) are supplemented with 10-minute binned data (black markers), while the solid red lines represent the median post-fit models.}
    \label{fig:grid_fits}
\end{figure*}

We processed the remaining candidates through the same multi-step pipeline. The phase-folded light curves and post-fit transit models for the six newly validated planets---including the two inner companions in the TOI-2392 system---are shown in Figure \ref{fig:grid_fits}. The complete set of derived physical properties, orbital ephemerides, and TSM values for the seven planets is given in Table \ref{tab:params_master}.

\subsubsection{The Three-Planet System TOI-2392}

Beyond TOI-2392.03 (Section 6.1), we validated the two inner companions, giving a compact three-planet system. TOI-2392.02 orbits at 0.049 AU ($P = 3.41$ days) and has an equilibrium temperature of $T_{\mathrm{eq}} \approx 1657$ K. Its radius of $R_p = 3.07_{-0.17}^{+0.17}\,R_\oplus$ and inferred mass of $M_p = 9.8_{-4.3}^{+7.7}\,M_\oplus$ give a bulk density of $\rho_p \approx 1871\,\mathrm{kg\,m^{-3}}$. Without a dynamical mass, we describe it only as a hot, sub-Neptune-sized planet.

Its neighbor, TOI-2392.01, orbits at 0.071 AU ($P = 5.91$ days). It is larger, with a radius of $R_p = 3.71_{-0.21}^{+0.20}\,R_\oplus$ and an inferred bulk density of $\rho_p \approx 1456\,\mathrm{kg\,m^{-3}}$. The period ratios of the system are $P_{01}/P_{02} \approx 1.74$ and $P_{03}/P_{01} \approx 1.55$. We do not claim any mean-motion resonance for these ratios.

\subsubsection{Physical Properties of the Remaining Systems}

The four single-candidate systems, TOI-1277.02, TOI-1737.01, TOI-4450.01, and TOI-7721.01, have radii between $3.08$ and $7.19\,R_\oplus$, above the radius valley that separates rocky super-Earths from larger planets \citep{Fulton2017}. The mass-radius relation assigns them low bulk densities (Table \ref{tab:params_master}); as the masses are inferred from the radii, this is an expected output of the relation rather than independent evidence for gaseous envelopes, which would need to be tested with RV masses.

TOI-7721.01 is the largest planet of the sample ($R_p = 7.19_{-0.44}^{+0.44}\,R_\oplus$), in a 2.56-day orbit. Its inferred bulk density is $\rho_p \approx 644\,\mathrm{kg\,m^{-3}}$ and its atmospheric scale height is $H_{\mathrm{atm}} \approx 529$ km. TOI-1277.02 orbits an active host on a 14.85-day orbit. Its radius is smaller ($R_p = 3.08_{-0.16}^{+0.17}\,R_\oplus$), but its low surface gravity and small host star ($R_* = 0.86\,R_\odot$) give a TSM of 52.4.

\begin{table*}[ht]
\centering
\caption{Derived Orbital and Physical Parameters for the Seven Validated Exoplanets}
\label{tab:params_master}
\renewcommand{\arraystretch}{1.3} 
\resizebox{\textwidth}{!}{%
\begin{tabular}{lccccccccc}
\toprule
\textbf{Target} & \textbf{$P$ (days)} & \textbf{$T_0$ (BJD)} & \textbf{$a$ (AU)} & \textbf{$b$} & \textbf{$R_p$ ($R_\oplus$)} & \textbf{$M_p$ ($M_\oplus$)$^a$} & \textbf{$\rho_p$ ($\mathrm{kg\,m^{-3}}$)} & \textbf{$T_{\mathrm{eq}}$ (K)$^b$} & \textbf{TSM$^b$} \\
\midrule
TOI-1277.02 & $14.85593_{-0.00002}^{+0.00002}$ & $2458684.9711_{-0.0018}^{+0.0019}$ & $0.1153_{-0.0077}^{+0.0071}$ & $0.20_{-0.14}^{+0.16}$ & $3.08_{-0.16}^{+0.17}$ & $9.6_{-4.1}^{+7.9}$ & $1826_{-783}^{+1437}$ & $666_{-18}^{+21}$ & $52.4$ \\
TOI-1737.01 & $14.44801_{-0.00002}^{+0.00002}$ & $2458717.0130_{-0.0012}^{+0.0011}$ & $0.1334_{-0.0103}^{+0.0093}$ & $0.52_{-0.09}^{+0.08}$ & $3.25_{-0.14}^{+0.14}$ & $10.6_{-4.6}^{+8.4}$ & $1703_{-743}^{+1310}$ & $1124_{-37}^{+44}$ & $29.0$ \\
TOI-2392.01 & $5.912364_{-0.000006}^{+0.000006}$ & $2459037.8670_{-0.0012}^{+0.0013}$ & $0.0705_{-0.0059}^{+0.0056}$ & $0.37_{-0.19}^{+0.14}$ & $3.71_{-0.21}^{+0.20}$ & $13.4_{-5.9}^{+10.5}$ & $1456_{-650}^{+1110}$ & $1375_{-47}^{+58}$ & $22.9$ \\
TOI-2392.02 & $3.407218_{-0.000006}^{+0.000008}$ & $2459039.4265_{-0.0042}^{+0.0021}$ & $0.0485_{-0.0040}^{+0.0038}$ & $0.41_{-0.18}^{+0.13}$ & $3.07_{-0.17}^{+0.17}$ & $9.8_{-4.3}^{+7.7}$ & $1871_{-821}^{+1435}$ & $1657_{-56}^{+69}$ & $21.4$ \\
TOI-2392.03 & $9.169453_{-0.000048}^{+0.000034}$ & $2459038.6294_{-0.0033}^{+0.0040}$ & $0.0937_{-0.0079}^{+0.0076}$ & $0.45_{-0.17}^{+0.13}$ & $2.44_{-0.17}^{+0.16}$ & $6.6_{-2.8}^{+5.2}$ & $2561_{-1085}^{+1990}$ & $1194_{-43}^{+51}$ & $11.3$ \\
TOI-4450.01 & $10.69247_{-0.00002}^{+0.00002}$ & $2459396.3329_{-0.0014}^{+0.0013}$ & $0.0987_{-0.0077}^{+0.0080}$ & $0.61_{-0.07}^{+0.06}$ & $4.29_{-0.24}^{+0.25}$ & $17.1_{-7.8}^{+13.5}$ & $1180_{-515}^{+927}$ & $879_{-32}^{+35}$ & $35.8$ \\
TOI-7721.01 & $2.56430_{-0.00006}^{+0.00006}$ & $2461046.3558_{-0.0011}^{+0.0010}$ & $0.0372_{-0.0030}^{+0.0032}$ & $0.54_{-0.09}^{+0.08}$ & $7.19_{-0.44}^{+0.44}$ & $42.7_{-19.8}^{+43.8}$ & $644_{-297}^{+639}$ & $1207_{-44}^{+47}$ & $63.8$ \\
\bottomrule
\multicolumn{10}{l}{\footnotesize $^a$ Planetary masses are not dynamically measured but are statistically inferred from the radius using the \texttt{mr\_forecast} empirical relations, assuming $e=0$.} \\
\multicolumn{10}{l}{\footnotesize $^b$ $T_{\mathrm{eq}}$ assumes a Bond albedo $A_B \approx 0.3$ and the TSM uses this $T_{\mathrm{eq}}$ (see Section 5.2); the TSM is evaluated at the median values and does not include the uncertainty of the inferred mass.} \\
\end{tabular}%
}
\end{table*}

\clearpage

\section{Discussion}

\subsection{Demographic Context}

\begin{figure*}[ht]
    \centering
    \includegraphics[width=\textwidth]{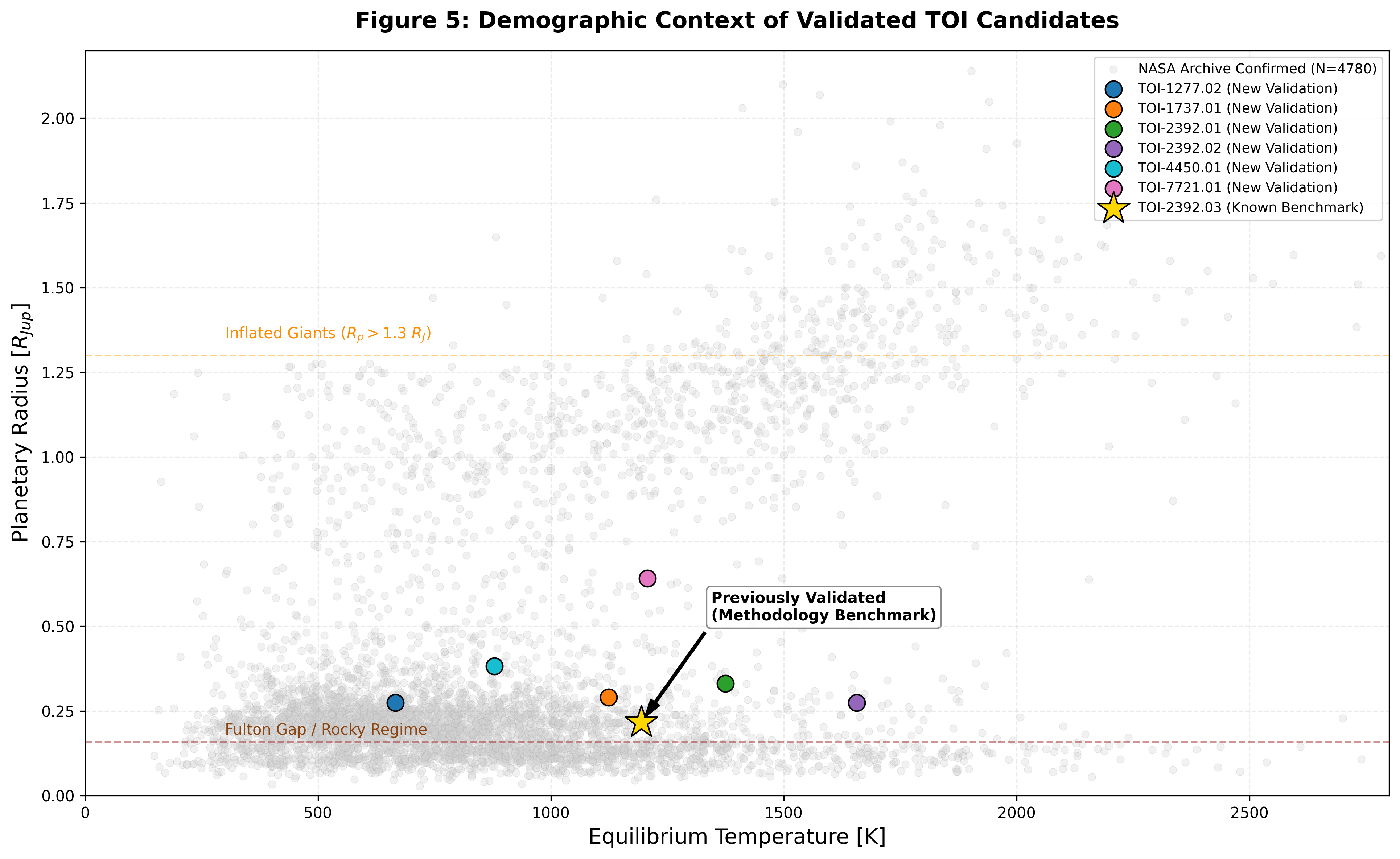}
    \caption{Planetary radius (in Jupiter radii) versus equilibrium temperature (K) for our validated TESS candidates, superimposed on the confirmed exoplanet population from the NASA Exoplanet Archive ($N=4780$, faint grey circles). The red dashed line denotes the approximate upper boundary of the rocky terrestrial regime (the Fulton Gap). All seven targets, including TOI-2392.03 (yellow star), lie above this boundary. Our $T_{\mathrm{eq}}$ assume $A_B \approx 0.3$, whereas archive values follow heterogeneous conventions (often zero albedo), so the comparison is approximate.}
    \label{fig:demographics}
\end{figure*}

To place our systems in context, we overlay the derived planetary radii and equilibrium temperatures on the confirmed population from the NASA Exoplanet Archive (Figure \ref{fig:demographics}). All seven planets, from the coolest TOI-1277.02 ($666$ K) to the hottest TOI-2392.02 ($1657$ K), lie above the radius valley that separates rocky super-Earths from larger planets \citep{Fulton2017}. TOI-7721.01 ($P = 2.56$ d, $R_p = 7.19\,R_\oplus$) and TOI-2392.02 ($P = 3.41$ d, $R_p = 3.07\,R_\oplus$) lie within the region commonly adopted for the Neptunian desert ($P < 4$ d, $3 < R_p < 8\,R_\oplus$; e.g., as defined by \citealt{Collier2026}), although TOI-2392.02 is close to its lower radius edge.

Because the masses used here are inferred from the radii, we do not draw conclusions on the envelope content of these planets. If RV follow-up confirms low bulk densities for the hotter planets ($T_{\mathrm{eq}} \gtrsim 1000$ K; the TOI-2392 planets, TOI-1737.01, and TOI-7721.01), these planets would be useful tests of atmospheric mass-loss models, since XUV photoevaporation and core-powered mass loss predict that such planets lose their primordial H/He envelopes unless they have massive cores. The planets validated here are therefore natural targets for RV mass measurements.

\subsection{Prospects for Atmospheric Spectroscopy with JWST and Ariel}

The Transmission Spectroscopy Metric (TSM) of \citet{Kempton2018} provides a standardized scaling of the expected SNR of near-infrared transmission spectroscopy features. As discussed in Section 5.2, our TSM values use $A_B \approx 0.3$ and an inferred mass, and should be considered as relative rankings. None of our planets reaches the TSM threshold of 90 suggested by \citet{Kempton2018} for high-quality targets with $1.5 < R_p < 10\,R_\oplus$. Within our sample, TOI-7721.01 and TOI-1277.02 have the highest values:
\begin{itemize}
    \item \textbf{TOI-7721.01 ($\mathrm{TSM} = 63.8$):} This planet ($R_p \approx 7.2\,R_\oplus$, $T_{\mathrm{eq}} \approx 1207$ K) has a large estimated scale height ($H_{\mathrm{atm}} \approx 529$ km) and low surface gravity, which favor transmission spectroscopy with JWST NIRISS/NIRSpec, if its atmosphere is not obscured by clouds or hazes. Since its validation is marginal ($\mathrm{FPP} = 0.0137$), additional vetting (e.g., high-resolution imaging and RV observations) would be advisable before investing JWST time.
    \item \textbf{TOI-1277.02 ($\mathrm{TSM} = 52.4$):} Although smaller, this planet benefits from a favorable planet-to-star radius ratio and a host star that is bright in the near-infrared ($J = 9.14$ mag). Its equilibrium temperature ($666$ K) is the lowest of the sample, which could place it in a different chemical regime than the hotter planets, where methane is expected to contribute more to the spectrum.
\end{itemize}

The three planets of TOI-2392 share the same host star, which would allow comparative atmospheric studies of planets with the same stellar irradiation history, although their TSM values (11.3--22.9) are lower than those of the two targets above.

\clearpage

\section{Conclusions}

In this study, we performed a pixel-level and Bayesian validation of seven TESS exoplanet candidates. Our framework combines automated signal retrieval, Gaia-based spatial contamination constraints, and \texttt{TRICERATOPS} false positive probabilities, followed by Bayesian transit modeling. Our conclusions are as follows:

\begin{enumerate}
    \item \textbf{Consistency check:} Processing TOI-2392.03, previously validated in the literature, with the same pipeline, we obtain period, epoch, and radius consistent with the published values (differences of $1.1\sigma$, $0.6\sigma$, and $0.3\sigma$) and an FPP of the same order as the published one. This checks our parameter extraction and gives one independent FPP comparison for this planet, but does not by itself calibrate the false-positive rejection of the pipeline.
    \item \textbf{Six newly validated exoplanets:} With the thresholds $\mathrm{FPP} < 0.015$ and $\mathrm{NFPP} < 0.001$, supported by centroid and Gaia contamination checks, we statistically validate six new planets with \texttt{TRICERATOPS}. TOI-7721.01 ($\mathrm{FPP} = 0.0137$) is close to the threshold. Their radii (3.1--7.2\,$R_\oplus$) place them above the radius valley; their compositions remain to be tested with RV masses.
    \item \textbf{A compact three-planet system:} TOI-2392 hosts three planets with radii of 2.4--3.7\,$R_\oplus$ and periods of 3.41, 5.91, and 9.17 days, with period ratios of 1.74 ($P_{01}/P_{02}$) and 1.55 ($P_{03}/P_{01}$).
    \item \textbf{Targets for atmospheric follow-up:} TOI-7721.01 and TOI-1277.02 have the highest TSM values in our sample (63.8 and 52.4), below the suggested threshold of \citet{Kempton2018}, and are of interest for JWST and \textit{Ariel} once their masses are measured and, for TOI-7721.01, its validation is reinforced.
\end{enumerate}

\section*{Data and Code Availability}
The TESS photometric data underlying this article are publicly available in the Mikulski Archive for Space Telescopes (MAST) at \url{https://mast.stsci.edu}. The analysis code, environment specifications, and derived results are publicly available at \url{https://github.com/biesro/TESS-Multi-Exoplanet-Validation}. An archival version of the software is available on Zenodo at \url{https://doi.org/10.5281/zenodo.24348805}.

\section*{Acknowledgments}
This paper includes data collected by the TESS mission, which are publicly available from the Mikulski Archive for Space Telescopes (MAST). Funding for the TESS mission is provided by NASA's Science Mission Directorate. We acknowledge the use of public TESS data from pipelines at the TESS Science Office and at the TESS Science Processing Operations Center.

This work has made use of data from the European Space Agency (ESA) mission \textit{Gaia} (\url{https://www.cosmos.esa.int/gaia}), processed by the \textit{Gaia} Data Processing and Analysis Consortium (DPAC, \url{https://www.cosmos.esa.int/web/gaia/dpac/consortium}). Funding for the DPAC has been provided by national institutions, in particular the institutions participating in the \textit{Gaia} Multilateral Agreement.

This research has made use of the NASA Exoplanet Archive and the Exoplanet Follow-up Observation Program (ExoFOP) website, which are operated by the California Institute of Technology, under contract with the National Aeronautics and Space Administration under the Exoplanet Exploration Program. This work makes use of an ingress observation of TOI-1277.02 obtained with the Las Cumbres Observatory Teide 1.0 m telescope and made available through ExoFOP-TESS; we thank Rick Schwarz and the TESS Follow-up Observing Program (TFOP) community for these data and their analysis.

The author thanks the developers of the open-source software used in this work. This research was carried out independently, without institutional funding.

Thanks to Dr.\ Angelica Psaridi for her availability and willingness to discuss the research.

\vspace{1em}
\facilities{TESS (MAST), Gaia, LCO}

\software{
    Astropy \citep{Astropy2022},
    batman \citep{Kreidberg2015},
    \texttt{dynesty} \citep{Speagle2020},
    Forecaster/\texttt{mr\_forecast} \citep{Chen2017},
    \texttt{juliet} \citep{Espinoza2019},
    Lightkurve \citep{Lightkurve2018},
    Matplotlib \citep{Hunter2007},
    \texttt{TRICERATOPS} \citep{Giacalone2021}
}

\clearpage

\appendix

\section{Spatial Contamination Maps}
This appendix provides the spatial contamination maps for the remaining four host stars in our sample (TOI-1277, TOI-1737, TOI-4450, and TOI-7721). Similar to Figure \ref{fig:sc_2392}, these figures show the TESS Target Pixel Files alongside the Gaia DR3 maps. They illustrate the local stellar environment and the proper-motion-corrected neighboring sources relative to the SPOC pipeline apertures.

\begin{figure*}[htbp]
    \centering
    \vspace*{1cm}
    \includegraphics[width=0.85\textwidth]{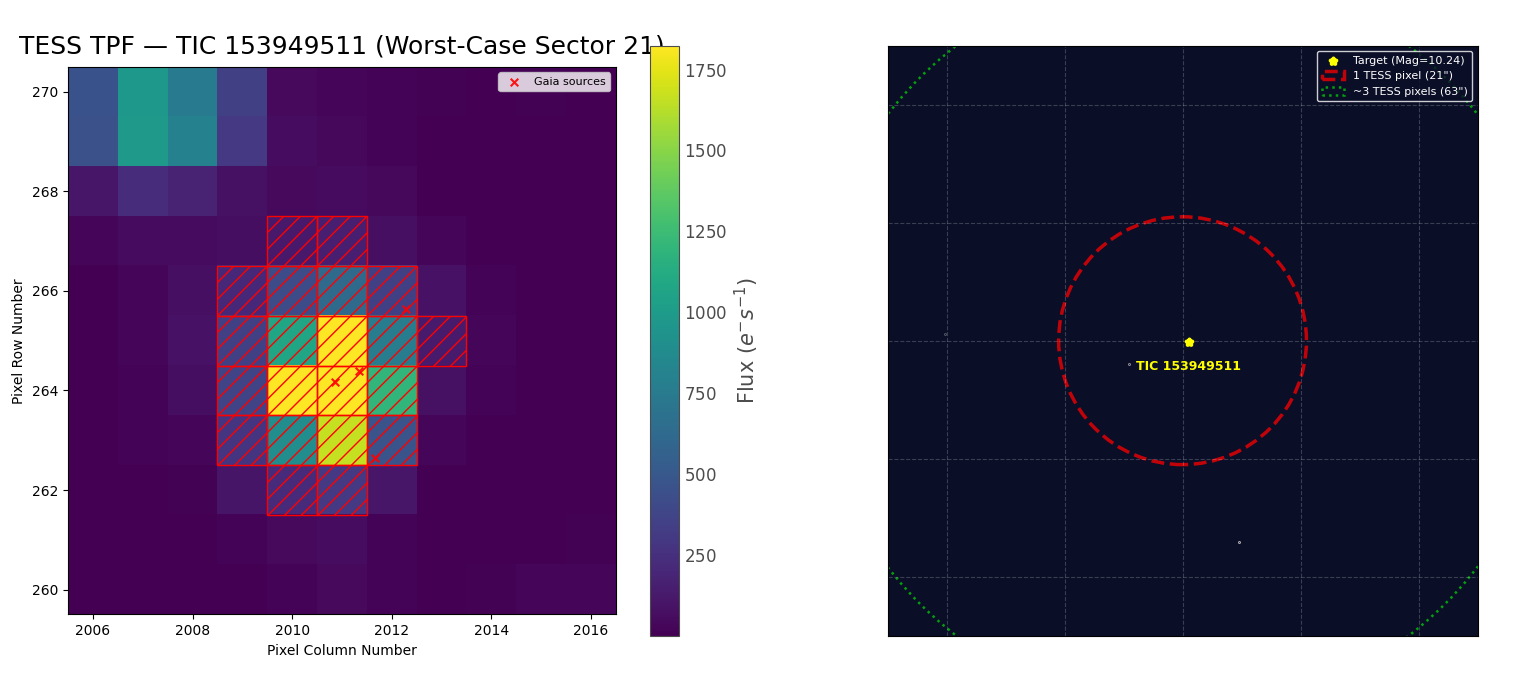}
    \caption{Spatial contamination analysis for TOI-1277. \textit{Left}: TESS Target Pixel File overlaid with the SPOC pipeline aperture mask. \textit{Right}: Gaia DR3 map ($1\arcmin$ search radius).}
    \label{fig:sc_1277}
\end{figure*}

\begin{figure*}[htbp]
    \centering
    \includegraphics[width=0.85\textwidth]{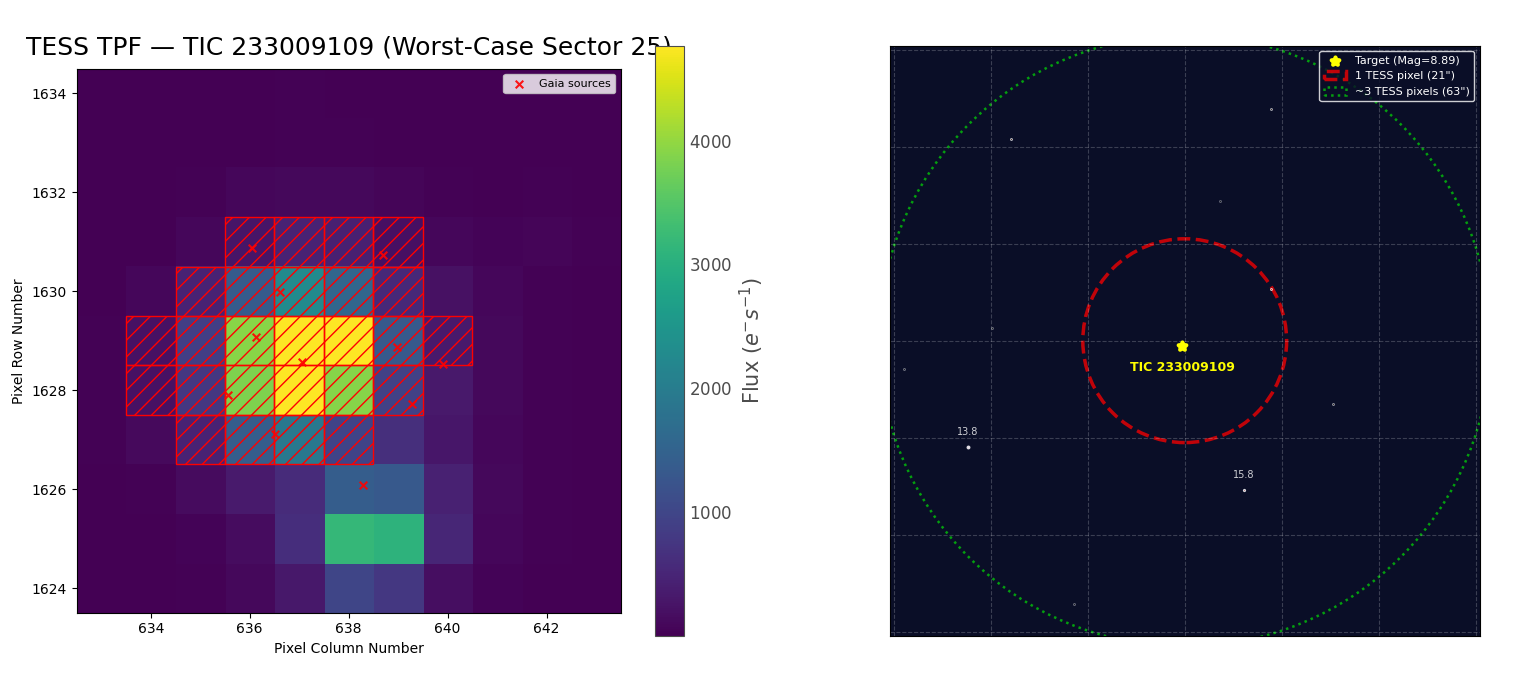}
    \caption{Spatial contamination analysis for TOI-1737. \textit{Left}: TESS Target Pixel File overlaid with the SPOC pipeline aperture mask. \textit{Right}: Gaia DR3 map ($1\arcmin$ search radius).}
    \label{fig:sc_1737}
\end{figure*}

\begin{figure*}[htbp]
    \centering
    \vspace*{3cm}
    \includegraphics[width=0.85\textwidth]{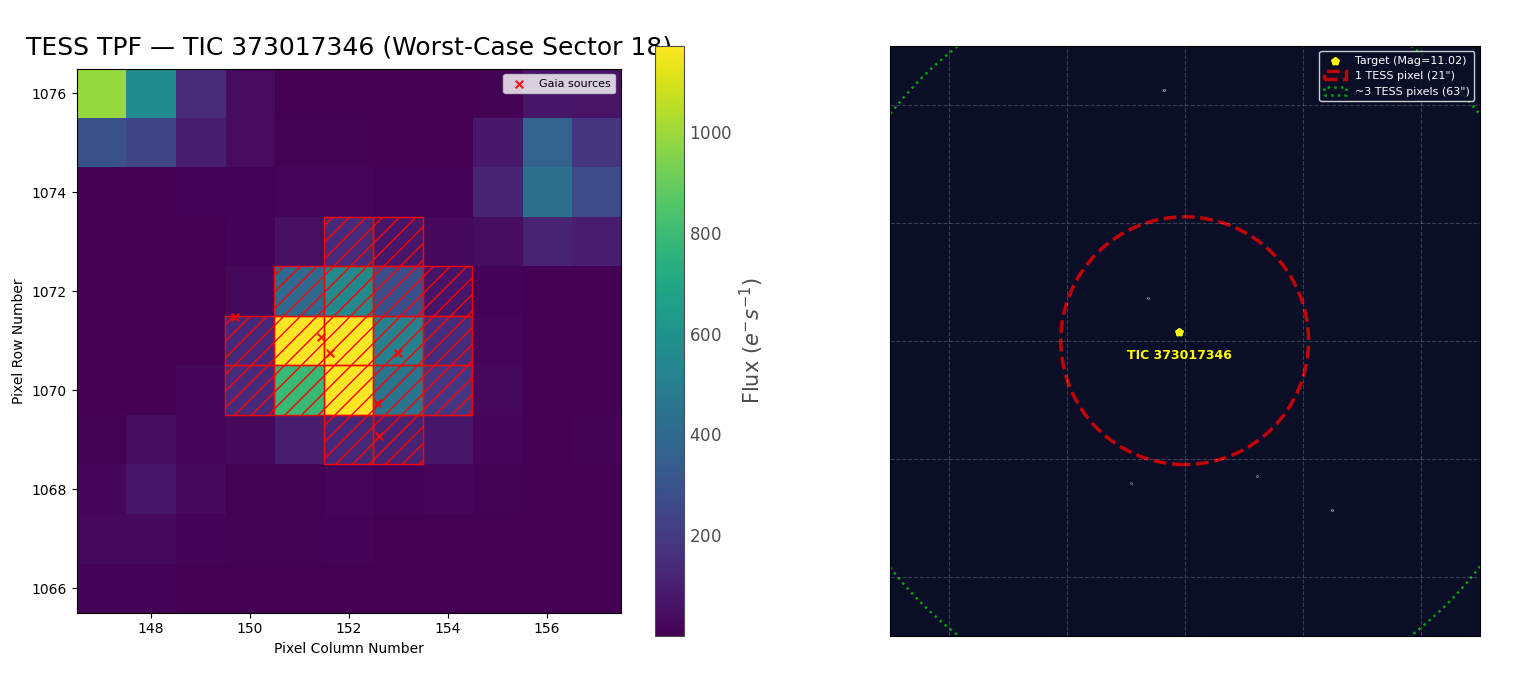}
    \caption{Spatial contamination analysis for TOI-4450. \textit{Left}: TESS Target Pixel File overlaid with the SPOC pipeline aperture mask. \textit{Right}: Gaia DR3 map ($1\arcmin$ search radius).}
    \label{fig:sc_4450}
\end{figure*}

\begin{figure*}[htbp]
    \centering
    \vspace*{1cm}
    \includegraphics[width=0.85\textwidth]{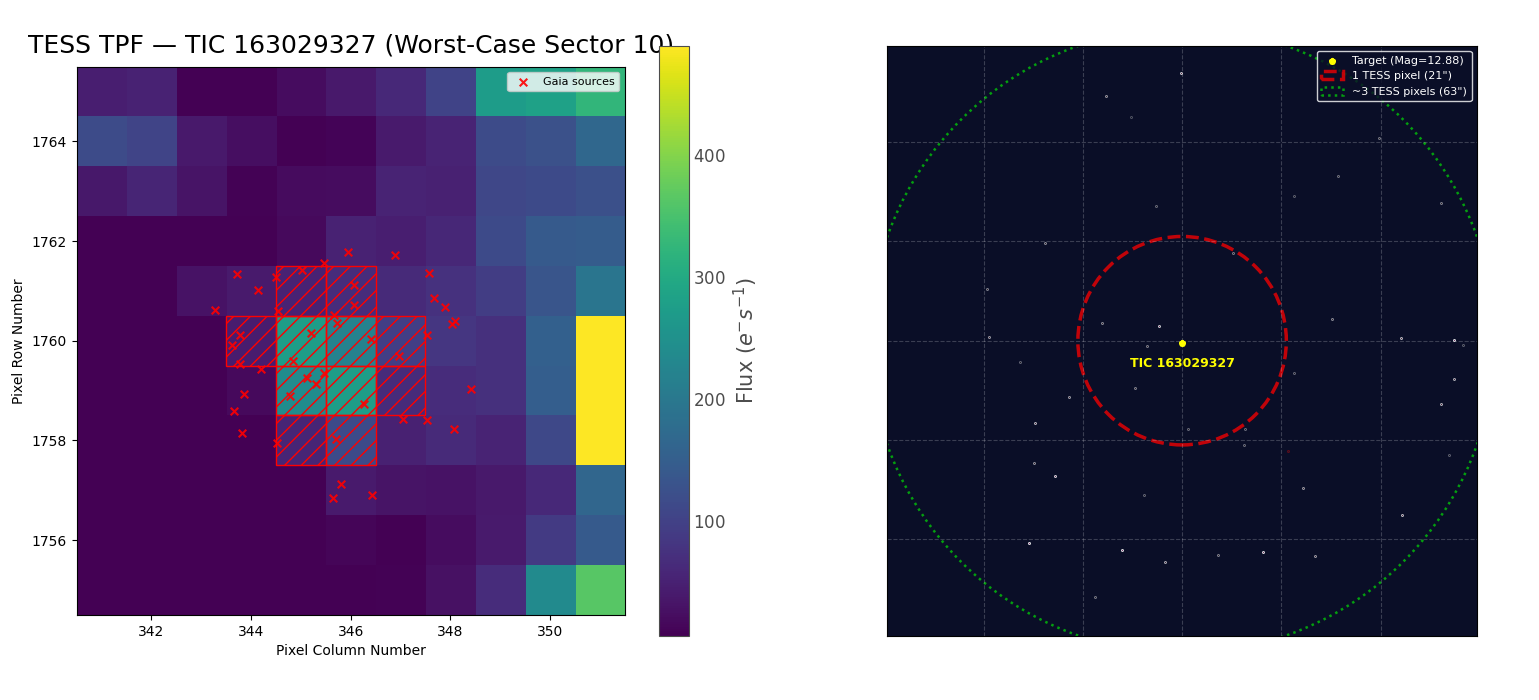}
    \caption{Spatial contamination analysis for TOI-7721. \textit{Left}: TESS Target Pixel File overlaid with the SPOC pipeline aperture mask. \textit{Right}: Gaia DR3 map ($1\arcmin$ search radius).}
    \label{fig:sc_7721}
\end{figure*}

\clearpage

\section{Derived Physical Posterior Distributions}
This appendix presents the Monte Carlo posterior distributions for the physical and orbital parameters of the six newly validated exoplanets, derived by combining the \texttt{juliet} nested sampling chains with the \texttt{mr\_forecast} empirical relation. The format is the same as in Figure \ref{fig:posteriors_2392}.

\begin{figure*}[htbp]
    \centering
    \vspace*{1cm}
    \includegraphics[width=0.65\textwidth]{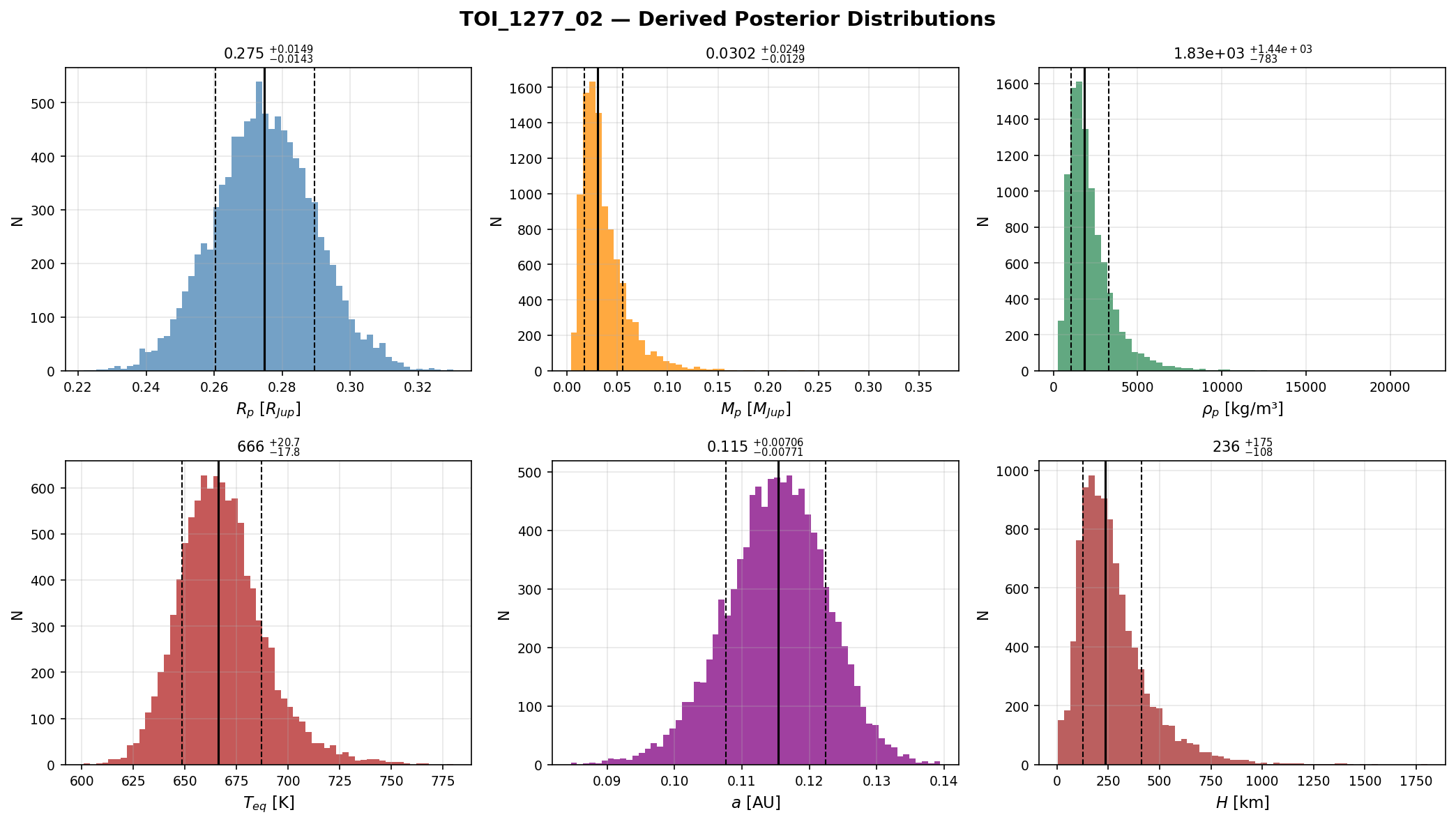}
    \caption{Derived physical posterior distributions for TOI-1277.02. Vertical lines denote the median (solid) and $1\sigma$ (dashed) confidence intervals.}
    \label{fig:post_1277_02}
\end{figure*}

\begin{figure*}[htbp]
    \centering
    \vspace*{2cm}
    \includegraphics[width=0.65\textwidth]{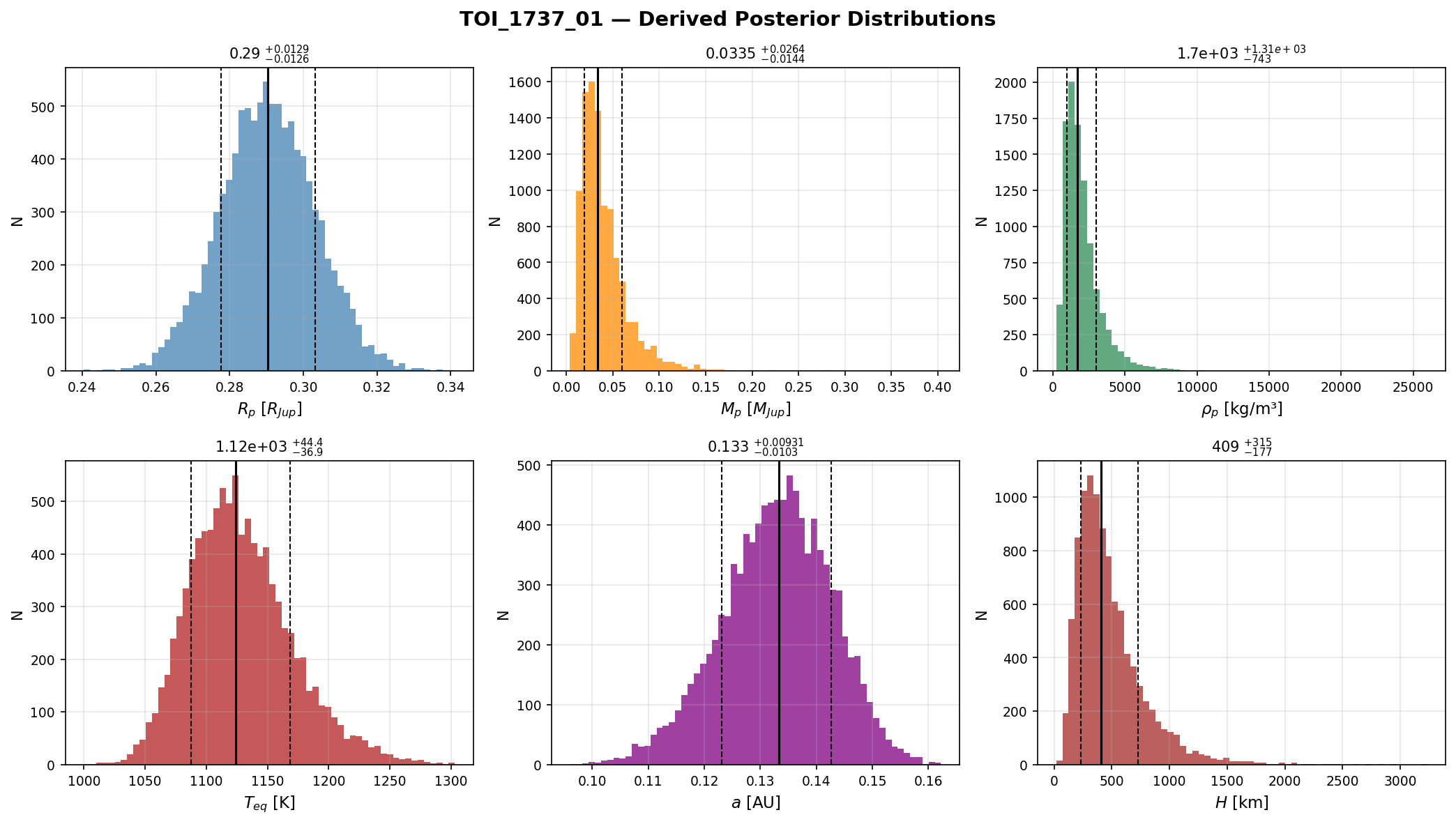}
    \caption{Derived physical posterior distributions for TOI-1737.01. Vertical lines denote the median (solid) and $1\sigma$ (dashed) confidence intervals.}
    \label{fig:post_1737_01}
\end{figure*}

\begin{figure*}[htbp]
    \centering
    \vspace*{3cm}
    \includegraphics[width=0.65\textwidth]{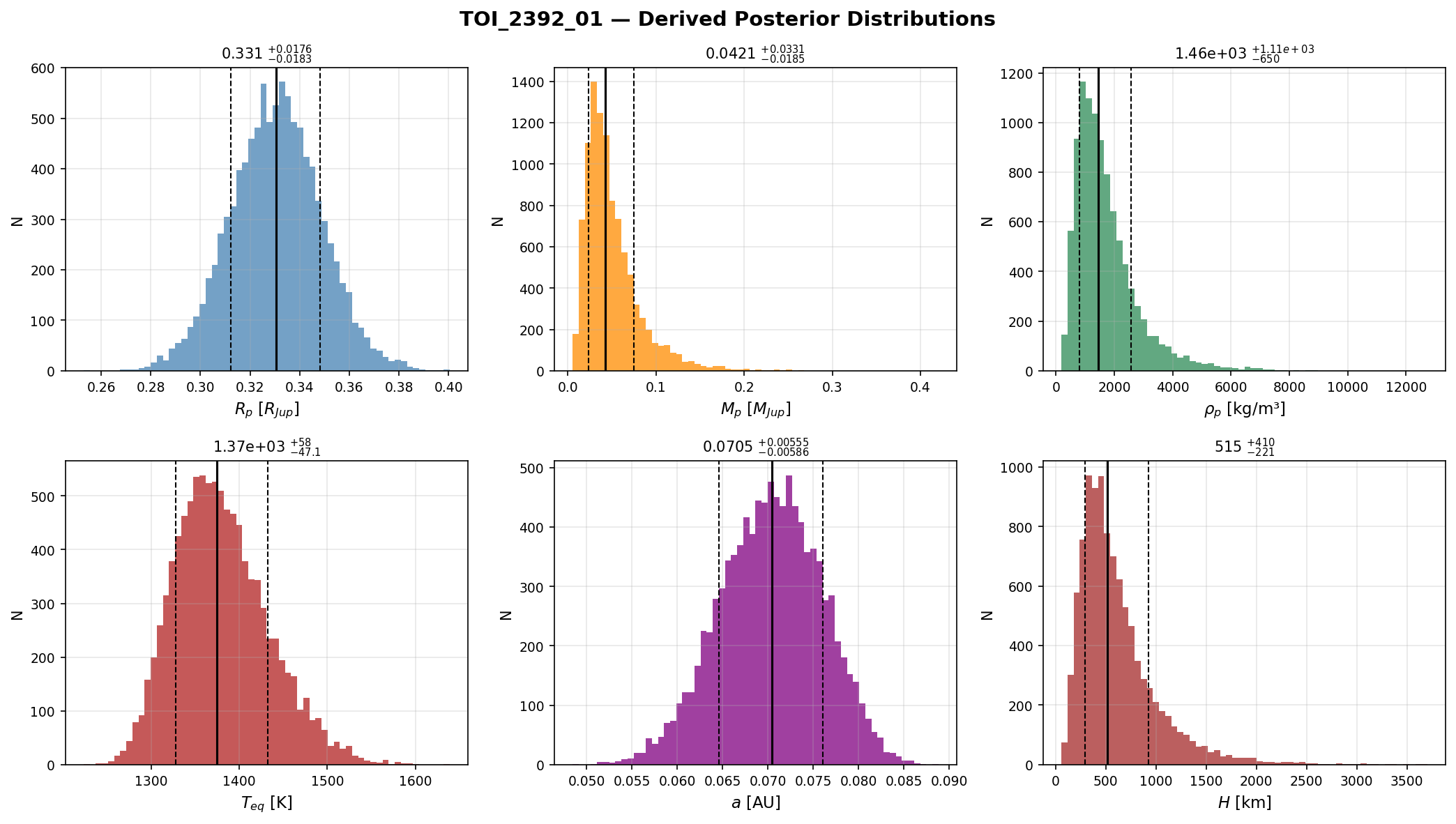}
    \caption{Derived physical posterior distributions for TOI-2392.01. Vertical lines denote the median (solid) and $1\sigma$ (dashed) confidence intervals.}
    \label{fig:post_2392_01}
\end{figure*}

\begin{figure*}[htbp]
    \centering
    \includegraphics[width=0.65\textwidth]{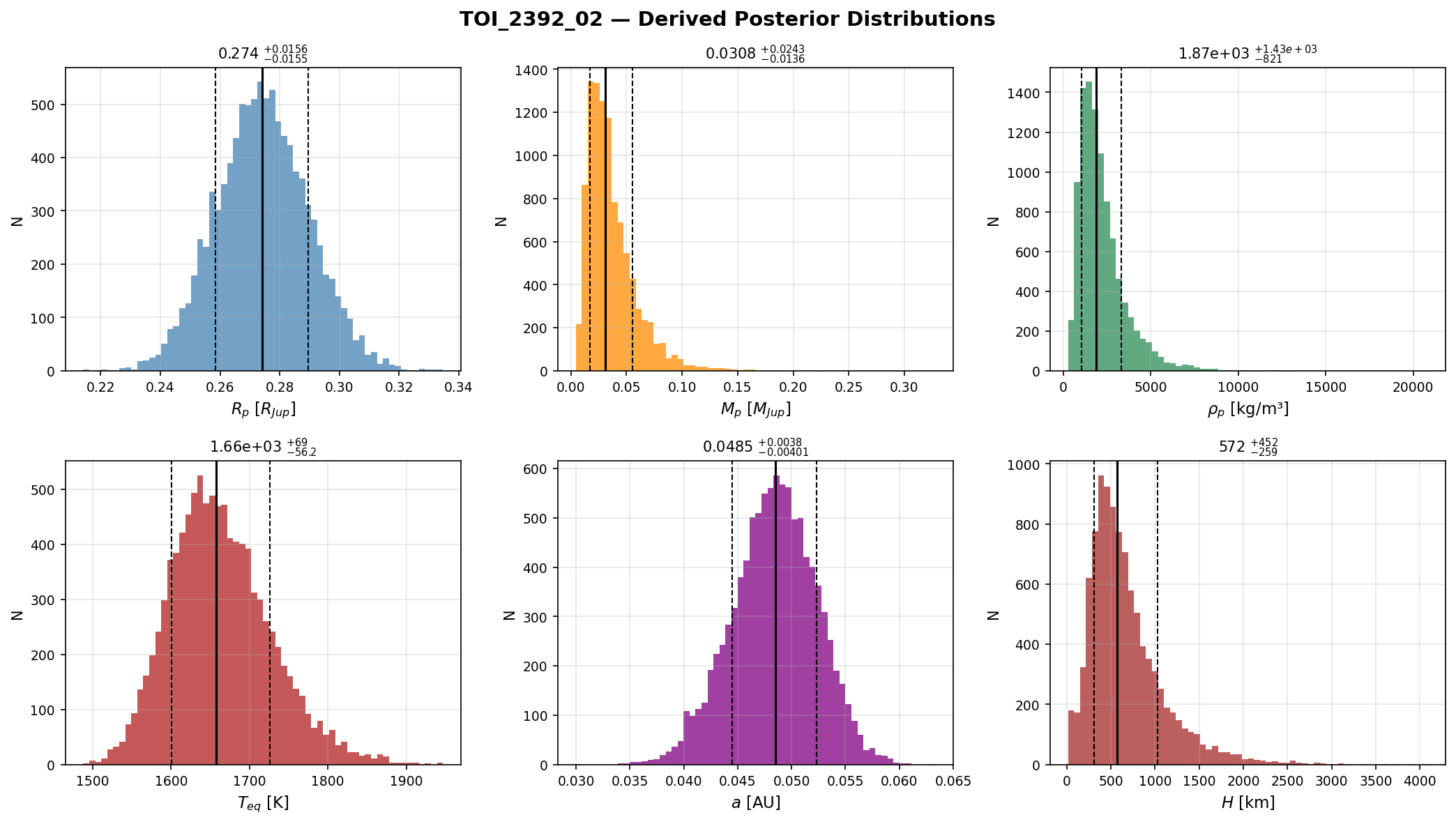}
    \caption{Derived physical posterior distributions for TOI-2392.02. Vertical lines denote the median (solid) and $1\sigma$ (dashed) confidence intervals.}
    \label{fig:post_2392_02}
\end{figure*}

\begin{figure*}[htbp]
    \centering
    \vspace*{3cm}
    \includegraphics[width=0.65\textwidth]{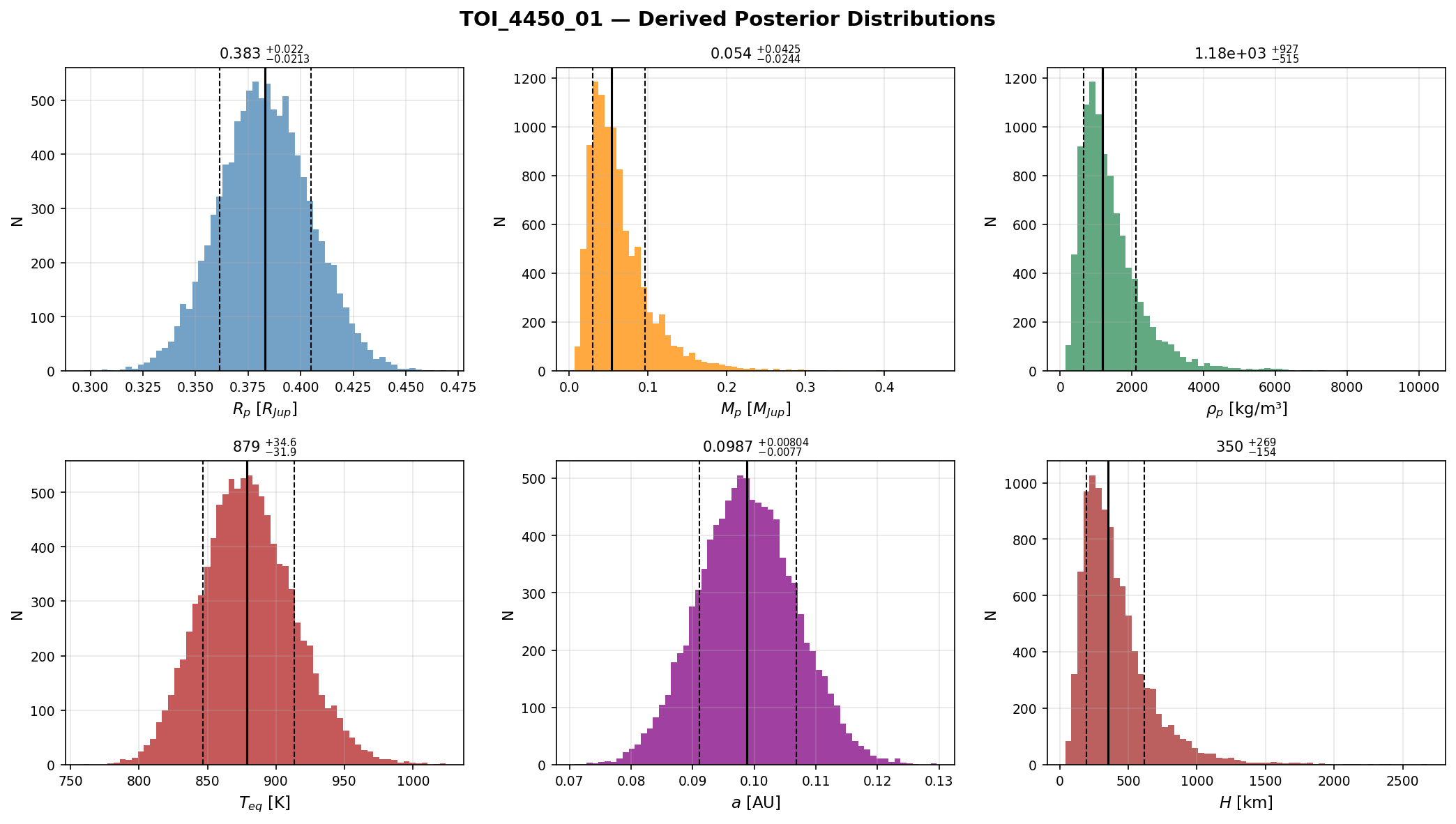}
    \caption{Derived physical posterior distributions for TOI-4450.01. Vertical lines denote the median (solid) and $1\sigma$ (dashed) confidence intervals.}
    \label{fig:post_4450_01}
\end{figure*}

\begin{figure*}[htbp]
    \centering
    \includegraphics[width=0.65\textwidth]{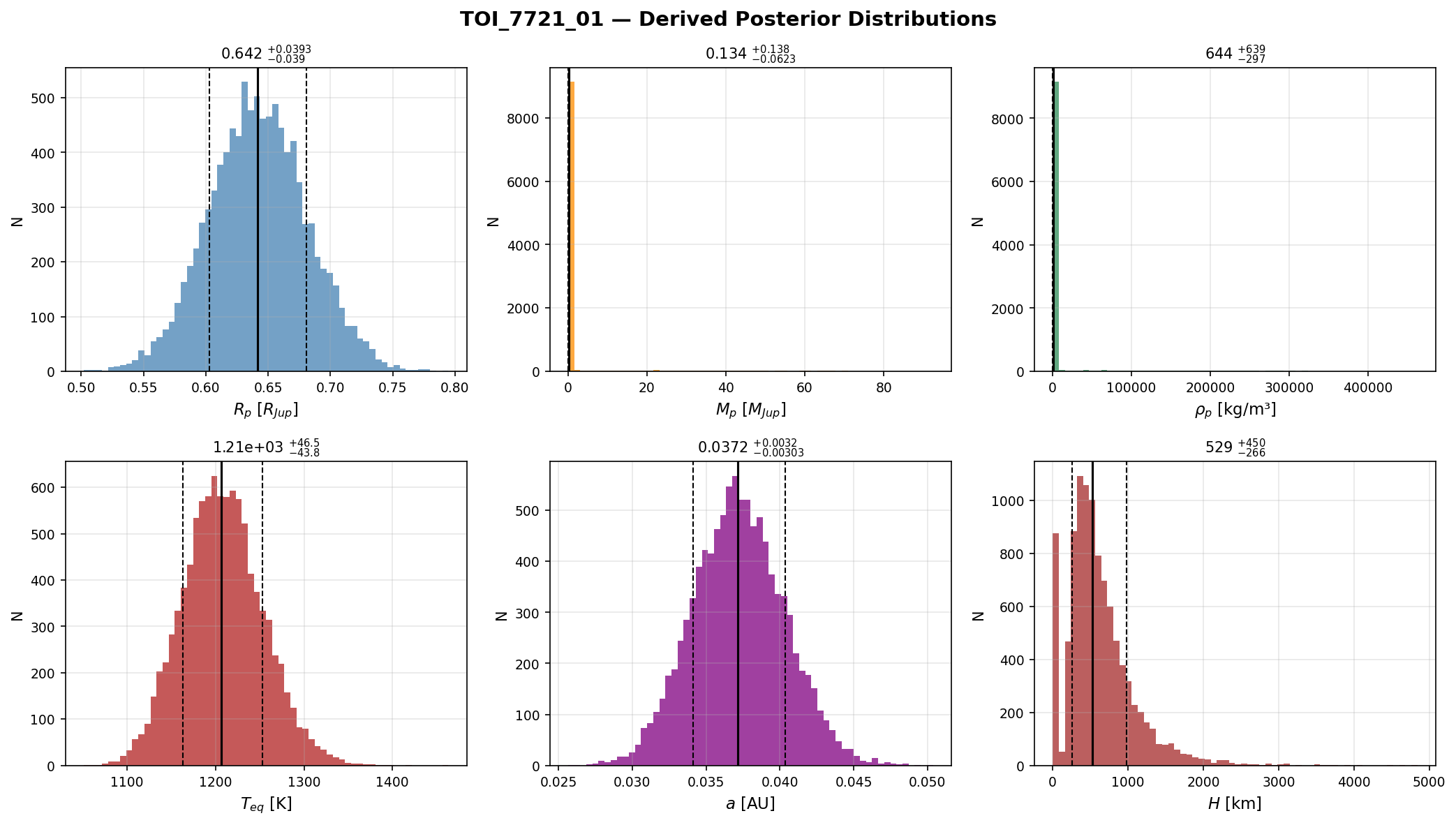}
    \caption{Derived physical posterior distributions for TOI-7721.01. Vertical lines denote the median (solid) and $1\sigma$ (dashed) confidence intervals.}
    \label{fig:post_7721_01}
\end{figure*}


\clearpage

\begin{thebibliography}{}
\bibitem[Astropy Collaboration et al.(2022)]{Astropy2022} Astropy Collaboration, Price-Whelan, A. M., Lim, P. L., et al. 2022, \apj, 935, 167
\bibitem[Chen \& Kipping(2017)]{Chen2017} Chen, J., \& Kipping, D. 2017, \apj, 834, 17
\bibitem[Collier et al.(2026)]{Collier2026} Collier, M., Giacalone, S., Derfer, B., et al. 2026, AJ, in press (arXiv:2608.21560)
\bibitem[Espinoza et al.(2019)]{Espinoza2019} Espinoza, N., Kossakowski, D., \& Brahm, R. 2019, \mnras, 490, 2262
\bibitem[Fulton et al.(2017)]{Fulton2017} Fulton, B. J., Petigura, E. A., Howard, A. W., et al. 2017, \aj, 154, 109
\bibitem[Gaia Collaboration et al.(2023)]{Gaia2023} Gaia Collaboration, Vallenari, A., Brown, A. G. A., et al. 2023, \aap, 674, A1
\bibitem[Giacalone et al.(2021)]{Giacalone2021} Giacalone, S., Dressing, C. D., Jensen, E. L. N., et al. 2021, \aj, 161, 24
\bibitem[Guerrero et al.(2021)]{Guerrero2021} Guerrero, N. M., Seager, S., Huang, C. X., et al. 2021, \apjs, 254, 39
\bibitem[Hadjigeorghiou et al.(2025)]{Hadjigeorghiou2025} Hadjigeorghiou, A., Armstrong, D. J., Cui, K., et al. 2025, arXiv e-prints, arXiv:2509.17645
\bibitem[Hunter(2007)]{Hunter2007} Hunter, J. D. 2007, Computing in Science \& Engineering, 9, 90
\bibitem[Kempton et al.(2018)]{Kempton2018} Kempton, E. M.-R., Bean, J. L., Louie, D. R., et al. 2018, \pasp, 130, 114401
\bibitem[Kov{\'a}cs et al.(2002)]{Kovacs2002} Kov{\'a}cs, G., Zucker, S., \& Mazeh, T. 2002, \aap, 391, 369
\bibitem[Kreidberg(2015)]{Kreidberg2015} Kreidberg, L. 2015, \pasp, 127, 1161
\bibitem[Lafarga et al.(2026)]{Lafarga2026} Lafarga, M., Armstrong, D. J., Cui, K., et al. 2026, \mnras, 548, stag512
\bibitem[Lightkurve Collaboration et al.(2018)]{Lightkurve2018} Lightkurve Collaboration, Cardoso, J. V. d. M., Hedges, C., et al. 2018, Astrophysics Source Code Library, ascl:1812.013
\bibitem[Lissauer et al.(2012)]{Lissauer2012} Lissauer, J. J., Marcy, G. W., Rowe, J. F., et al. 2012, \apj, 750, 112
\bibitem[Morton(2012)]{Morton2012} Morton, T. D. 2012, \apj, 761, 6
\bibitem[Ricker et al.(2015)]{Ricker2015} Ricker, G. R., Winn, J. N., Vanderspek, R., et al. 2015, JATIS, 1, 014003
\bibitem[Savanov(2022)]{Savanov2022} Savanov, I. S. 2022, Astrophysics, 65, 75
\bibitem[Speagle(2020)]{Speagle2020} Speagle, J. S. 2020, \mnras, 493, 3132
\bibitem[Stassun et al.(2019)]{Stassun2019} Stassun, K. G., Oelkers, R. J., Paegert, M., et al. 2019, \aj, 158, 138
\bibitem[Torres et al.(2011)]{Torres2011} Torres, G., Fressin, F., Batalha, N. M., et al. 2011, \apj, 727, 24
\end{thebibliography}
\end{document}